%% file: n1068-usero.tex
\documentclass{aa}
	 
\usepackage{graphicx}  
\usepackage{txfonts}

\newcommand{\dala}[2]{$#1\times10^{#2}$}
\begin{document}

  \title{Molecular Gas Chemistry in AGN}

\authorrunning{A. Usero et al.}

  \subtitle{I. The IRAM 30m Survey of NGC~1068}

\author{A. Usero\inst{1,2}, S. Garc\'{\i}a-Burillo\inst{1}, A. Fuente\inst{1}, and J.
Mart\'{\i}n-Pintado\inst{3}, { N. J. Rodr\'{\i}guez-Fern\'andez\inst{4}}}

\offprints{A. Usero,\\
\email{antonio.u@imaff.cfmac.csic.es}}

\institute{Observatorio Astron\'omico Nacional (OAN), C/ Alfonso XII 3, 
  28014 Madrid, Spain \and
Instituto de Matem\'aticas y F\'{\i}sica Fundamental, CSIC, C/ Serrano 113bis, 28006 Madrid, Spain
 \and Instituto de Estructura de la
  Materia, DAMIR-CSIC, C/ Serrano 121, 28006
  Madrid, Spain \and  { LERMA (UMR 8112), 
Observatoire de Paris, 61, Av. de l'Observatoire, 75014 Paris, France}}  

\date{Received 1 December 2003 / Accepted 5 February 2004}

\abstract{There is observational evidence that nuclear winds and X-rays can  heavily influence
the physical conditions and chemical abundances of  molecular gas in the circumnuclear disks (CND)
of Active Galactic Nuclei  (AGN). In this paper we probe the chemical status of molecular gas in the
CND  of \object{NGC~1068}, a prototypical  Seyfert 2 galaxy. Precedent claims that the chemistry of
molecular gas in the nucleus of  \object{NGC~1068} is {\sl abnormal} by galactic standards were
based on the high  HCN/CO luminosity ratio measured in the CND. Results from new observations
obtained in this survey have { served to derive}  abundances of molecular species such as SiO, CN,
HCO$^+$,  HOC$^+$, H$^{13}$CO$^+$ and HCO. These  estimates are complemented by a re-evaluation of
molecular abundances for  HCN, CS and CO, based on previously published single-dish and
interferometer  observations of \object{NGC~1068}. We report on the first detection of   SiO
emission in the CND of \object{NGC~1068}. The estimated large  abundance of SiO in the CND,
X(SiO)$\sim$(5-10)$\times$10$^{-9}$, cannot be  attributed to shocks related to star formation, as
there is little evidence  of a recent starburst in the nucleus of \object{NGC~1068}. Alternatively,
we  propose that silicon chemistry is driven by intense X-ray processing of  molecular gas. We also
report on the first extragalactic detection of the  reactive ion HOC$^+$. Most remarkably, the
estimated HCO$^+$/HOC$^+$  abundance ratio in the nucleus of \object{NGC~1068}, $\sim$30--80, is the
smallest ever measured in molecular gas. The abundances derived for all  molecules that have been
the subject of this survey are compared with the  predictions of models invoking either
oxygen-depletion or X-ray chemistry in  molecular gas. Our conclusions favour an overall scenario
where the CND of  \object{NGC~1068} has become a giant X-ray Dominated Region (XDR).
 
 } 

\maketitle 

\section{Introduction} Active Galactic Nuclei (AGN) are able to inject vast 
amounts of energy into their { host galaxies}, carried by strong radiation fields 
and rapidly moving jets. It is predictable that AGN should have a disruptive 
influence on the gas reservoir near their central engines. There is 
multi-wavelength observational evidence that the general properties of neutral 
interstellar matter in AGN differ from { those} of quiescent star-forming disks 
and starburst galaxies (Genzel et al. \cite{genz98}; Laurent et al. 
\cite{laur00}). In particular, molecular gas close to the central engines of 
active galaxies can be exposed to a strong X-ray irradiation. While the 
accretion disks of AGN are strong UV emitters, the bulk of the UV flux can be 
attenuated by neutral gas column densities of only 
N(H)$\sim$10$^{21}$cm$^{-2}$. Hard X-ray photons (2--10~keV) can penetrate neutral gas column 
densities out to N(H)$\sim$10$^{23}$-10$^{24}$cm$^{-2}$, however. Therefore, 
X-ray dominated regions (XDR) could become the dominant sources of emission 
for molecular gas in the harsh environment of circumnuclear disks (CND) of 
AGN, as originally argued by Maloney et al. (\cite{mall96}).

\begin{figure}[!htp]

\resizebox{\hsize}{!} { \includegraphics{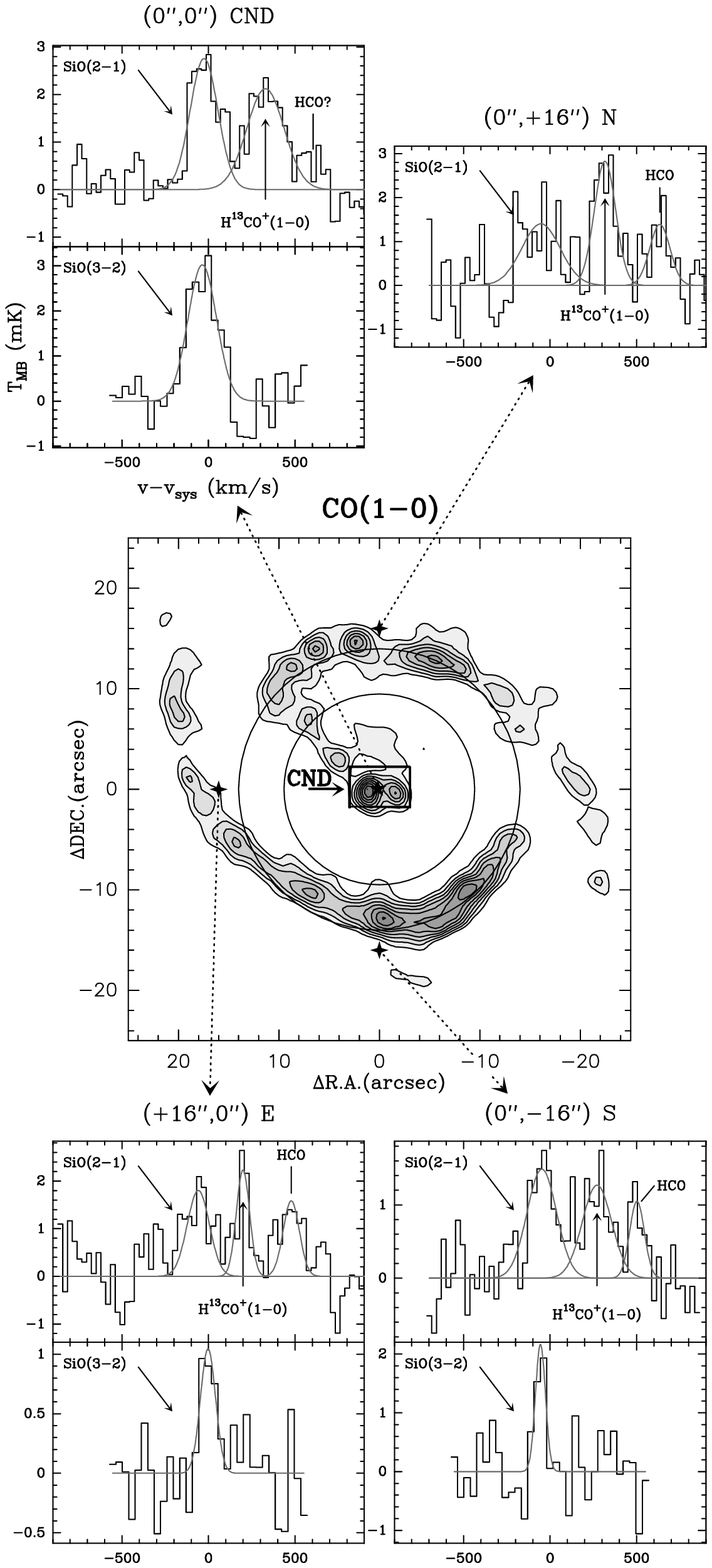}}
 \caption{Emission spectra of the 2--1 and 3--2 lines of SiO detected in the 
inner 3~kpc of \object{NGC~1068}. Four starred markers, overlaid on the 
CO(1--0) integrated intensity map of Schinnerer et al. 
(\cite{schi00}), highlight the { central positions of the beams} in the disk where we searched for 
SiO emission: the central offset (0\arcsec,0\arcsec) coincides with the 
position of the AGN, given by the S1 
compact radio-source of Gallimore et al. (\cite{gall96b}) 
($\alpha_{2000}$=02$^{\mathrm{h}}$42$^{\mathrm{m}}$40$^{\mathrm{s}}$.71,$\delta_{2000}$=
--00\degr00\arcmin47.9\arcsec), while offsets N, S and E probe the SiO emission over the 
starburst ring. Emission in the H$^{13}$CO$^+$(1--0) and HCO(1--0) lines is 
detected in the CND and over the starburst ring. The circles represent the 
beam sizes at 130.3~GHz (19\arcsec) and 86.8~GHz (28\arcsec).} \label{PosFig} 
\end{figure}

First observational evidence that the physical and chemical properties of 
molecular gas in the CND of AGN depart from `normality' came from the 
single-dish and interferometer observations of HCN and CO emission in 
\object{NGC~1068} (Tacconi et al. \cite{tacc94}; Sternberg et al. 
\cite{ster94}). This prototypical Seyfert 2 galaxy hosts a circumnuclear 
starburst ring of $\sim$2.5-3\,kpc--diameter (see Fig.~\ref{PosFig}); the 
ring delimits a 2.3~kpc stellar bar detected by Scoville et al. 
(\cite{scov88}) in the NIR. The strong emission detected in the 1--0 and 2--1 
CO lines coming from the starburst ring corroborates that massive star 
formation is fed by a significant gas reservoir (Planesas et al. \cite{plan89},
\cite{plan91}; Helfer
\& Blitz \cite{helf95}; Schinnerer et al.  \cite{schi00}).
 Significant CO emission arises also from a 200\,pc CND of 
M(H$_2$)$\sim$5$\times$10$^{7}$M$_{\sun}$ { (inferred using a
N(H$_2$)/I(CO) conversion factor of 2.2$\times$10$^{20}$~cm$^{-2}$~(K~km~s$^{-1}$)$^{-1}$, from
Solomon \& Barrett \cite{solo91})}.  The CND, partly
resolved into two  knots, surrounds the position of the active nucleus identified as the compact 
radio-source S1 in the map of Gallimore et al. (\cite{gall96a}). Most 
remarkably, the CND is prominent in HCN emission (Tacconi et al. 
\cite{tacc94}). According to the analysis of Sternberg et al. 
(\cite{ster94}), the high HCN/CO intensity ratio measured by Tacconi et al. 
(\cite{tacc94}) ($\sim$1--10) leads to an abnormally high HCN/CO abundance 
ratio in the nucleus of \object{NGC~1068}: N(HCN)/N(CO)$\sim$a few~10$^{-3}$-10$^{-2}$, i.e., 
the highest ratio ever found in the centre of any galaxy.

Different explanations have been advanced to quantify the possible link 
between the anomalous HCN chemistry and the presence of an active nucleus in 
\object{NGC~1068}. The selective depletion of gas-phase oxygen in the dense 
molecular clouds would explain the high HCN-to-CO abundance ratio (Sternberg et al. \cite{ster94};
Shalabiea \&  Greenberg \cite{shal96}). The same 
oxygen depletion scheme predicts a lower-than-normal abundance of all 
oxygen-bearing species. Alternatively, an 
increased X-ray ionization of molecular clouds near the AGN could enhance the 
abundance of HCN (Lepp \& Dalgarno \cite{lepp96}). Furthermore, X-rays could 
evaporate small ($\sim$10~\AA) silicate grains, increasing the fraction in 
gas phase of all refractory elements and subsequently enhancing the abundance 
of some molecules (e.g., SiO) in X-ray irradiated molecular gas (Voit 
\cite{voit91}; Mart\'{\i}n-Pintado et al. \cite{mart00}). While the 
aforementioned scenarios { succeed} to reproduce the measured enhancement of HCN 
relative to CO in \object{NGC~1068}, their predictions about the abundances 
of other molecular species differ significantly. The lack of tight 
observational constraints for these models, prompted by the first 
mm-observations made in \object{NGC~1068}, has hampered thus far the choice 
of an optimum scenario, however.

In this paper we discuss the results of a molecular survey made in 
\object{NGC~1068} with the IRAM 30m mm-telescope. \object{NGC~1068} is the optimum target to {
quantify} the feed-back of  activity and star formation on the chemistry of molecular
gas. Furthermore,  the spatial resolution of the 30m telescope is well suited to discern between
the emission coming from the star forming ring and that coming from the CND.  We discuss the results
obtained from new mm-observations of 6 molecular species. The list includes:  SiO(v=0, J=2--1 and
J=3--2), HCO(J=3/2--1/2, F=2--1),  H$^{13}$CO$^{+}$(J=1--0), HCO$^{+}$(J=1--0), HOC$^{+}$(J=1--0)
and  CN(N=2--1). For { comparison purposes}, we include in our analysis the  results from previous
single-dish and interferometer observations of CO  (J=1--0 and 2--1 from Schinnerer et
al. \cite{schi00}; J=4--3 from Tacconi et  al. \cite{tacc94}), HCN (J=1--0 and 4--3) (Tacconi et
al. \cite{tacc94}) and  CS (J=2--1) (Tacconi et al. \cite{tacc97}). This data { base} has served for
estimating the abundances of eight molecular species in the CND of  \object{NGC~1068} using LVG
model calculations. The inferred abundances are  compared with the predictions of models invoking
either oxygen-depletion or  X-ray chemistry in molecular gas.  We present in Section~\ref{sec2} the
30m observations made for this survey as  well as the data compiled from previous works on
\object{NGC~1068}.  Section~\ref{sec3} presents the results obtained from our SiO study and their
implications for the CND chemistry. Section~\ref{sec4} is devoted to discuss  the chemistry of the
HOC$^{+}$/HCO$^{+}$ active ions. The molecular gas  inventory of the CND is globally presented and
discussed in Section~\ref{sec5}.  We discuss in Section~\ref{sec6} the interpretation of these
results in the  framework of different chemistry models and summarize the main conclusions of  this
work in Section~\ref{sec7}.


\section{Observations} \label{sec2}

The observations have been carried out { in four sessions} from January 2000 to August 2002 with  
the IRAM 30m radiotelescope at Pico Veleta (Spain). We used 3 SIS receivers 
tuned in single-sideband mode in the 1~mm, 2~mm and 3~mm bands to observe 
several transitions of the molecular species shown in Tab.~\ref{ObsPar}, 
which summarizes the relevant parameters of these observations. We have 
obtained single-pointed spectra toward the nucleus of \object{NGC~1068} for 
all the molecules with the exception of SiO, HCO and H$^{13}$CO$^{+}$, for 
which we obtained partial maps by observing three additional positions on the 
starburst ring (see Sect.~\ref{sec3}). The line temperature scale used by 
default throughout the paper is T$_{\mathrm{MB}}$, i.e., main brightness 
temperature. T$_{\mathrm{MB}}$ is related to antenna temperature, 
T$_{\mathrm{A}}^*$, by T$_{\mathrm{A}}^*$=T$_{\mathrm{MB}}\times\eta_B$; the 
values assumed for $\eta_B$ are listed in Tab.~\ref{ObsPar}. When explicitly 
stated, T$_{\mathrm{MB}}$ temperatures are corrected by a source coupling 
factor, $f_{S}$\footnote{Correction for dilution: 
T$\rightarrow f_S$T with $f_S$=$\Omega_{\mathrm{beam}}/\Omega_{\mathrm{S}}$, where 
$\Omega_{\mathrm{beam}}$ is the beam area and $\Omega_{\mathrm{S}}$ the
 area of the emitting 
region estimated from the CO(1-0) interferometer map.
}; this factor accounts for the estimated dilution
of the  
source within the beam. To improve the stability of spectral baselines, the 
observations { have been carried out in beam-switching mode, with an azimuthal switch of
$\pm4\arcmin$ with a frequency of 0.5~Hz}. Only linear polynomials were used in the baseline 
correction.

In this paper we also use the data from previously published HCN, CS and CO 
observations of \object{NGC~1068} made with the IRAM Plateau de Bure 
Interferometer-PdBI (HCN(1--0): Tacconi et al. \cite{tacc94}; CS(2--1): 
Tacconi et al. \cite{tacc97}; CO(1--0) and CO(2--1): Schinnerer et al. 
\cite{schi00}). Complementary observations of high J transitions (J=4--3) of 
CO and HCN, taken at James Clerk Maxwell Telescope-JCMT (Tacconi et al. 
\cite{tacc94}), are also included. The main parameters of these observations 
are listed in Tab.~\ref{ObsPar}.

Hereafter, we will assume a distance to \object{NGC~1068} of 14.4~Mpc 
(Bland-Hawthorn et al. \cite{blan97}). This implies 1\arcsec=72~pc. The 
assumed heliocentric systemic velocity is v$_{\mathrm{sys}}$=1137~km~s$^{-1}$ 
(from NASA/IPAC Extragalactic Database (NED)).

\begin{table*}[!htp] \caption{Main parameters of the new 30m observations 
(top). { Typical receiver} and system temperatures are shown as T$_{rec}$ and T$_{sys}$, 
respectively. We also show the relevant parameters for previous observations 
used in this { work (bottom)}. See original references for details.}

\label{ObsPar} \input{n1068-usero-tab1.txt}

\end{table*}

\section{The IRAM 30m Survey of \object{NGC~1068}}
 \subsection{SiO Emission in \object{NGC~1068} } 
\label{sec3} 
\object{NGC~1068} was { originally part} of a larger 
extragalactic survey searching for SiO emission in starbursts (Usero et al. 
2003 in prep). Different mechanisms have been found thus far to explain the 
enhancement of SiO abundances in molecular gas in galaxies: either related to 
recent star formation (\object{NGC~253}: Garc\'{i}a-Burillo et al. 
\cite{buri00}), to the disruption of galaxy disks by large-scale shocks 
(\object{M~82}: Garc\'{i}a-Burillo et al. \cite{buri01}) or to the X-ray irradiation of molecular
clouds (Milky
Way: Mart\'in-Pintado et al. \cite{mart00}).

We show in Fig.~\ref{PosFig} the 4 positions over the \object{NGC~1068} disk where we 
searched for SiO emission. To better constrain the physical conditions of the 
gas, we have observed simultaneously the J=2--1 and J=3--2 rotational 
transitions of SiO. SiO(2--1) emission is detected at every offset, 
while SiO(3--2), { very} prominent in the CND, is detected in 2 out of the 3 
positions mapped over the ring. The observing grid { was chosen} to 
discriminate between SiO emission coming from the starburst ring 
(N[0\arcsec,+16\arcsec], E[16\arcsec,0\arcsec] and S[0\arcsec,--16\arcsec]) 
and that coming from the circumnuclear disk (CND[0\arcsec,0\arcsec]). 
Parameters of the gaussian fits to the lines detected are listed in 
Tab.~\ref{FitsTabA}.
\begin{table*}[!htp] 

 \caption {Parameters of gaussian fits to the 
SiO/H$^{13}$CO$^{+}$/HCO lines observed in \object{NGC~1068}. {  Errors (in brackets) are
1-$\sigma$. For the non-detection of SiO(3--2) in the N position we give a 3-$\sigma$ upper limit.}}

\label{FitsTabA} {\input{n1068-usero-tab2.txt}}

\end{table*}


\subsubsection{{Emission} in the Starburst Ring} \label{sec3.1}

These observations show that SiO emission is widespread in the starburst ring 
of \object{NGC~1068}. Where detected over the ring, SiO(3--2) lines are 
 narrower than SiO(2--1) lines. This result can be explained if, 
contrary to the compactness of SiO emission in the CND (see below), the 
emission of SiO on the ring extends significantly beyond a single SiO(3--2) 
beam. Within the errors, the I(SiO(3--2))/I(SiO(2--1)) integrated intensity 
ratios are $\sim$0.5 { in the two positions with detection of the 2mm line}.
 These ratios are slightly lowered to 
{ 0.4$\pm$0.1} if we apply a correction due to the different coupling factors of 
the 3--2 and 2--1 beams with the source { (correcting for dilution of the 
 nearly one-dimensional elongated arm inside the
beams, i.e., by a factor $\sim$19\arcsec/28$\arcsec$).}  

There are two precedents for the detection of large-scale SiO emission 
associated with ongoing { star formation:} \object{NGC~253} 
(Garc\'{i}a-Burillo et al. \cite{buri00}) and \object{M~82} 
(Garc\'{i}a-Burillo et al. \cite{buri01}). The derived enhancement of SiO 
abundances (X(SiO)$\sim$a few 10$^{-10}$--10$^{-9}$) takes place on scales of 
several hundred pc in these starbursts and has been interpreted as a 
signature of shocks driven by YSO, SN explosions and/or density waves. In the 
starburst ring of \object{NGC~1068}, a significant fraction of the stellar 
population ($\sim$40\% of the total optical light; Gonz\'alez-Delgado et al. 
\cite{gonz01}) has { typical ages $\leq$10$^{7}$~yr}. This supports that a 
recent short burst of star formation has occurred coevally throughout the 
ring on a time-scale of$\sim$10$^{6}$~yr { (Davies et al. \cite{davi98}).

Beside the detection} of the 1--0 line of H$^{13}$CO$^+$ (see 
Fig.~\ref{PosFig}), which is 
93~MHz redshifted with respect to the SiO(2--1) line,
 we have detected the emission of the strongest hyperfine 
component (F=2--1) of the J=3/2--1/2 line of HCO over the starburst ring. 
Observations of HCO in galactic clouds suggest that the abundance of this 
molecule is enhanced in Photon Dominated Regions (PDR). More recently, 
Garc\'{i}a-Burillo et al. (\cite{buri02}) have reported on the detection of 
widespread HCO emission in the nuclear starburst of \object{M~82}, where it 
traces the propagation of PDR chemistry in the disk.
Based on studies of HCO emission in Galactic PDR (Schenewerk, Snyder, \&
Hjalmarson \cite{sche86}; Schenewerk et al. \cite{sche88}), it is plausible to suppose that the HCO lines
should be optically thin also in the starburst ring of \object{NGC~1068}.
For H$^{13}$CO$^{+}$ we also consider optically thin emission and
the same excitation temperature as that assumed for HCO. These are reasonable guesses,
especially for T$_\mathrm{ex}$, as the two molecules have similar critical densities for the examined
transitions. In this case, the calculation of { the} HCO-to-H$^{13}$CO$^{+}$ column density ratio is
straightforward using the expression (Schenewerk et al. \cite{sche88}): 
\begin{equation}
\mathrm{
\frac{{ N(HCO)}}{N(H^{13}CO^+)}\simeq
{ \frac{12}{5}}\frac{I_{HCO}A_{HCO}^{-1}}{I_{H^{13}CO^+}A_{H^{13}CO^+}^{-1}}
}
\end{equation}

 where N is the total column density, I is the integrated intensity, and A is the Einstein
coefficient of the transition. We infer an average value for N(HCO)/N(H$^{13}$CO$^+$) of $\sim$8.
Adopting an average fractional abundance for H$^{13}$CO$^+$ of 10$^{-10}$ (Garc\'{\i}a-Burillo et
al. \cite{buri00}, \cite{buri01}), we derive X(HCO)$\sim$8$\times$10$^{-10}$.  The estimated
N(HCO)/(H$^{13}$CO$^+$) abundance ratios in prototypical PDR range from 30, in the \object{Orion
Bar}, to 3, in \object{NGC 7023} (Schilke et al. \cite{schi01}).

 Altogether, the 
detection of widespread SiO and HCO emission in the starburst ring of 
\object{NGC~1068} can be naturally explained by the chemical processing of 
molecular gas after a recent episode of star formation.

\subsubsection{SiO Emission in the Circumnuclear Disk (CND)} \label{sec3.2}

As  is shown in Fig.~\ref{PosFig}, the spatial resolution of the 30m in the 
3--2 line (19\arcsec) guarantees that the SiO(3--2) emission detected toward 
the CND has little if any contamination from the starburst ring (of 
$\sim$30$\arcsec$ diameter). The similar line-widths of the 2--1 and 3--2 SiO 
spectra at (0\arcsec,0\arcsec) provide further evidence that the bulk of the 
central SiO(3--2) emission comes from the CND. Furthermore, the linewidth of 
both SiO lines (FWZP=350~km~s$^{-1}$) coincides with the total line width of 
the CO(1--0) emission integrated within the CND, as derived from the 
interferometer map of Schinnerer et al. (\cite{schi00}). While the SiO(3--2) 
line at (0\arcsec,0\arcsec) has no significant contribution from the 
starburst ring, the situation is less clear in the case of the SiO(2--1) 
spectrum: the 28$\arcsec$ 30m beam at half power may { pick up} emission coming 
mostly from the southern ridge of the star forming ring (see Fig.~\ref{PosFig}). 
Taking into account that the SiO(2--1) line temperatures measured over the 
ring are a factor of 2 lower than in the CND, the derived upper limit for the 
`alien' contribution to the SiO(2--1) CND spectrum is $\sim$25$\%$, at most. 

The I(SiO(3--2))/I(SiO(2--1)) ratio in the CND is of
 { 0.7$\pm$0.1, once corrected for the contribution of the starburst ring
to the 2--1 CND line ($\times$1/0.75) and for the two-dimensional beam dilution of the CND
 ($\times$(28\arcsec/19\arcsec)$^2$).} { Simultaneously, we have evaluated the contribution of
 the CND to the SiO(2--1) spectra in the ring to be, at most, $\sim$25\%. 
 When we correct for this effect,
the I(SiO(3--2))/I(SiO(2--1)) average ratio on the ring derived in Sect.~\ref{sec3.1} is raised to
0.5$\pm$0.1, i.e., a factor 1.5 smaller than the ratio in the CND.
 Although the difference is only
marginal, it suggests that the excitation  of SiO lines in the CND is different from that of the
ring. In particular gas densities in the CND could be larger by a factor of $\sim$4 compared to the
starburst ring.}

A relevant contribution from the molecular bar to the SiO emission detected at 
(0\arcsec,0\arcsec) is also very unlikely for several reasons. First, the bar 
hardly stands out in the HCN and CS interferometer maps of 
\object{NCG~1068} (Tacconi et al. \cite{tacc97}): this is { a relevant result, 
as the} critical densities of HCN(1--0) and CS(2--1) lines are similar to that of 
SiO(1--0). Second, while weak CO emission is detected along the bar, it is 
significant only at v$<$v$_{\mathrm{sys}}$: this is at odds with the observed 
SiO line profiles, { roughly symmetric on both sides around v$_{\mathrm{sys}}$.

Most remarkably,} there is no evidence for significant recent star formation in the CND itself.
Several multiwavelength criteria have classified the nucleus of \object{NGC~1068} as a {\sl pure}
Seyfert nucleus, with little contribution from a nuclear starburst (MIR: Laurent et 
al. \cite{laur00}; NIR: Imanishi \cite{iman02}; Optical/Near-UV: 
Cid-Fernandes et al. \cite{cidf01}); the compact starburst emits $\leq1\%$ of 
the total IR luminosity (Marco \& Brooks \cite{marc03}). The circumnuclear 
stellar population is concentrated in a 50~pc core of `post-starburst' 
intermediate age stars (age $\geq$5-16$\times$10$^8$ yr) (Thatte et al. 
\cite{that97}).

We can exclude star formation either inside or outside the CND as the 
mechanism explaining the emission of SiO detected at (0\arcsec,0\arcsec). 
This poses the problem of the origin of SiO emission in the CND. The energy 
budget inside the CND seems to be largely dominated by the AGN itself; thus 
the chemistry of molecular gas, in particular the silicon chemistry, could be 
driven by non-stellar processes. We discuss in Section~\ref{SiOXdr} how the 
high abundances derived for SiO in the CND might be linked to the onset of 
XDR chemistry.


\subsection{Emission of Reactive Ions in NGC~1068: the HOC$^{+}$/HCO$^{+}$ 
Isomers} \label{sec4}

Detailed chemical models of XDR predict enhanced abundances of some reactive 
ions (e.g., H$_3^+$, HCO$^+$, SO$^{+}$ CO$^+$ and HCNH$^+$) as well as 
related neutral species (such as CN and HCN) (Maloney et al. \cite{mall96}; 
Black \cite{blac98a}, \cite{blac98b}; Lepp \& Dalgarno \cite{lepp96}). The 
tentative detection of CO$^{+}$ in the radio galaxy Cygnus A (Fuente et al. 
\cite{fuen00}) suggests that reactive ions may be used as an efficient 
diagnostic tool to study XDR chemistry in AGN. As part of this multi-species 
survey of \object{NGC~1068}, we have observed the 1--0 line of HCO$^+$ 
toward the CND. Most importantly, we have also searched for emission 
of its metastable isomer, HOC$^+$. There is 
recent observational evidence that X(HCO$^+$)/X(HOC$^+$) ratios, usually 
ranging from 300-6000 for dense molecular clouds in our Galaxy { (Apponi et al. 
\cite{appo97}, \cite{appo99}), can} reach values as low as 50-100 in 
UV-irradiated clouds (e.g., the prototypical PDR \object{NGC~7023}: Fuente et 
al. \cite{fuen03}). These results urged us to estimate the 
HCO$^+$--to--HOC$^+$ ratio in the X-ray bathed environment of an AGN.

\begin{figure}[!htp]

\centering \includegraphics[width=\hsize]{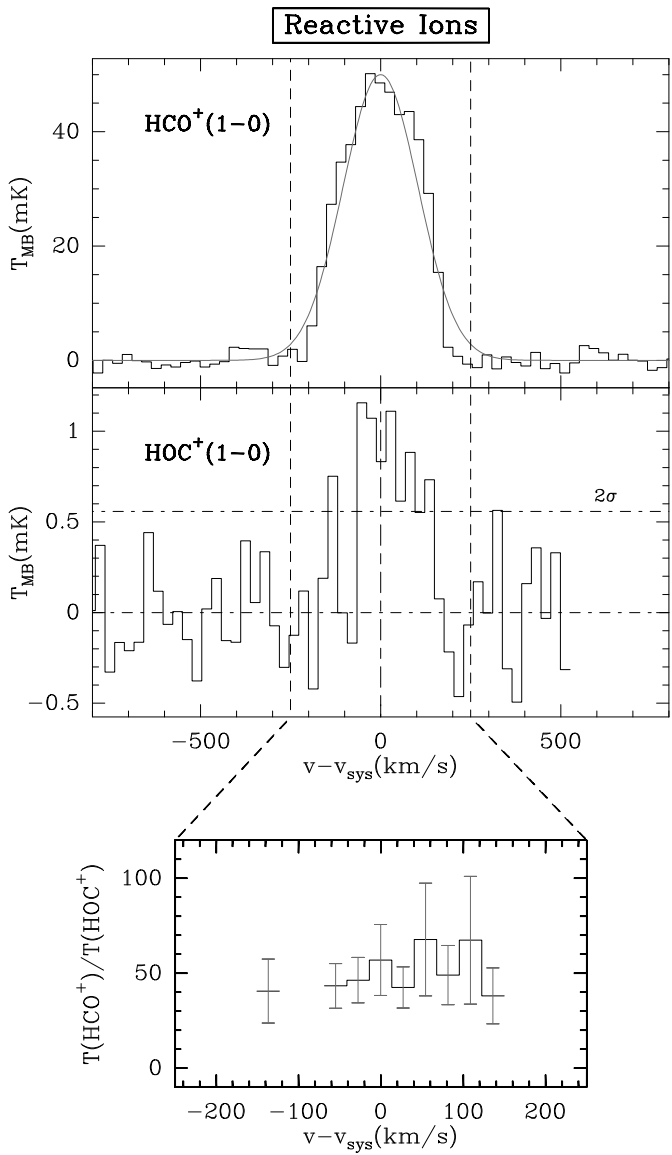} 
\caption{{\sl top} and {\sl middle}: HCO$^+$(1--0) and HOC$^+$(1--0) spectra 
of the CND of \object{NGC~1068}. {\sl bottom}: 
HCO$^+$(1--0)--to--HOC$^+$(1--0) temperature ratio profile derived for 
channels fulfilling T[HOC$^+$(1--0)]$>2\sigma$. { Error bars} are $\pm\sigma$.} \label{ReacSpec}
\end{figure}

Fig.~\ref{ReacSpec} shows the J=1--0 30m spectra of HCO$^+$ and HOC$^+$ 
observed toward the CND of \object{NGC~1068}. The emission of both species is 
detected. The interferometer HCO$^+$ map of Kohno et al. (\cite{kohn01}) 
shows that the CND largely dominates the emission of HCO$^{+}$ in the inner 3~kpc 
of \object{NGC~1068}. Moreover, we can estimate a conservative upper limit 
for the contribution of the starburst ring to the HCO$^+$ emission detected 
at (0\arcsec,0\arcsec). Following the same procedure used in 
Section~\ref{sec3.2}, here { adapted} to H$^{13}$CO$^+$, we estimate that 
$<$30\% of the (0\arcsec,0\arcsec) H$^{13}$CO$^+$ emission can be attributed 
to the starburst ring. We can reasonably extrapolate this estimate to 
HCO$^+$. Additional { evidence, similar to the one discussed in 
Section~\ref{sec3.2}, supports} that the 30m HCO$^+$ spectrum is heavily 
{ dominated} by emission coming from the CND.

The most remarkable result is the tentative detection of the HOC$^+$(1--0) 
line, the first thus far obtained in an external galaxy. HOC$^+$(1--0) 
emission is detected over 2$\sigma$ levels in a 215~km~s$^{-1}$ velocity 
range ($\sim$[--65~km~s$^{-1}$,+150~km~s$^{-1}$]). The emission integrated 
within this velocity window reaches a 8.5$\sigma$ significance level. The 
line profile of HOC$^+$ is noticeably asymmetrical with respect to 
v$_{\mathrm{sys}}$: HOC$^+$ emission is mostly detected at {\sl red} 
velocities. { As is shown} in Fig.~\ref{ReacSpec}, HCO$^+$--to--HOC$^+$ intensity 
ratios for v$\geq$v$_{\mathrm{sys}}$ range from $\sim$40 to 100. These 
surprisingly low { values rival the lowest} values thus far derived in 
PDR. The low HCO$^{+}$-to-HOC$^{+}$ intensity ratio measured in the CND of 
\object{NGC~1068} suggests that the chemistry of molecular gas could be 
driven by the pervading X/UV irradiation coming from the Seyfert 2 nucleus. 
Most remarkably, the asymmetry of the HOC$^+$ line profile suggests that 
whatever causes the enhancement of this active ion, the process responsible 
seems to be unevenly efficient inside the CND. As it is discussed in 
Section~\ref{HocXdr}, X-ray driven chemistry in the CND may 
{ satisfactorily explain} a dramatic change in the HCO$^{+}$-to-HOC$^{+}$ abundance ratio.

\subsection{CN Emission in \object{NGC~1068}}

 CN is a high-dipole radical typically found in dense regions ($\sim$10$^{5}$~cm$^{-3}$). The
abundance of CN is strongly linked to that of HCN. Theoretical models (Lepp \& Dalgarno
\cite{lepp96}) predict  large CN-to-HCN abundance ratios ($>$1) in XDR. The comparison of the CN and
HCN emission may thus provide a suitable diagnostic of the relevance of X-rays in the chemistry of
the CND.             

The CN(2--1) transition is split up into 18 hyperfine lines that appear blended into
three groups at frequencies $\sim$226.9~GHz, $\sim$226.7~GHz and $\sim$226.4~GHz.  We were able to
observe the two most intense groups of the transition (the 226.9~GHz and 226.7~GHz groups, hereafter
referred to as {\sl high frequency} and {\sl low frequency} respectively), although the {\sl low
frequency} group was only partially covered by the spectral bandwidth. The beam size at this {
frequency (11\arcsec) guarantees} that the detected CN(2--1) emission must be coming from the CND.     

The CN(2--1) spectrum is shown in Fig.~\ref{CndFigA} (main-beam temperature scaled to the CND; see
Section~\ref{profiles}). The measured high-frequency-to-low-frequency intensity ratio is below the
expected value for the optically thin limit { (high/low$\sim$1.64$\pm$0.14 instead of
1.80). However, this estimate is hampered by the insufficient baseline coverage in the spectrum.}

\section{Molecular Gas Inventory of the CND} 
\label{sec5}
 Understanding the 
peculiar chemistry of molecular gas revealed in the CND of \object{NGC~1068} 
requires a global analysis of its molecular inventory. Furthermore, higher 
spatial resolution is key to extracting the maximum information from the 30m 
spectra of SiO, HCO$^+$ and HOC$^+$ discussed above.

With this aim we have included in our analysis the information provided by 
published interferometer maps of \object{NGC~1068} obtained for CO, CS and 
HCN (Schinnerer et al. \cite{schi00}, Tacconi et al. \cite{tacc94}, 
\cite{tacc97}). These maps can help to improve our knowledge on the molecular 
abundances for species such as HCN or CS in the CND; due to their high 
spatial resolution, these observations are not hampered by source confusion 
between the CND itself and the starburst ring. In particular, the CO 
interferometer map allows us to estimate the molecular hydrogen column 
densities in the CND. Moreover, the spatio-kinematical information of the CO 
interferometer map is used for calculating the size and the location inside 
the CND of the gas components emitting at different velocities. Altogether, 
this information is employed in Sect.~\ref{SecLvg} to estimate via LVG models 
the abundances of several molecular species in the CND, separately, for the 
relevant velocity components.

\begin{figure}[!htp]

\centering 
\includegraphics[width=\hsize]{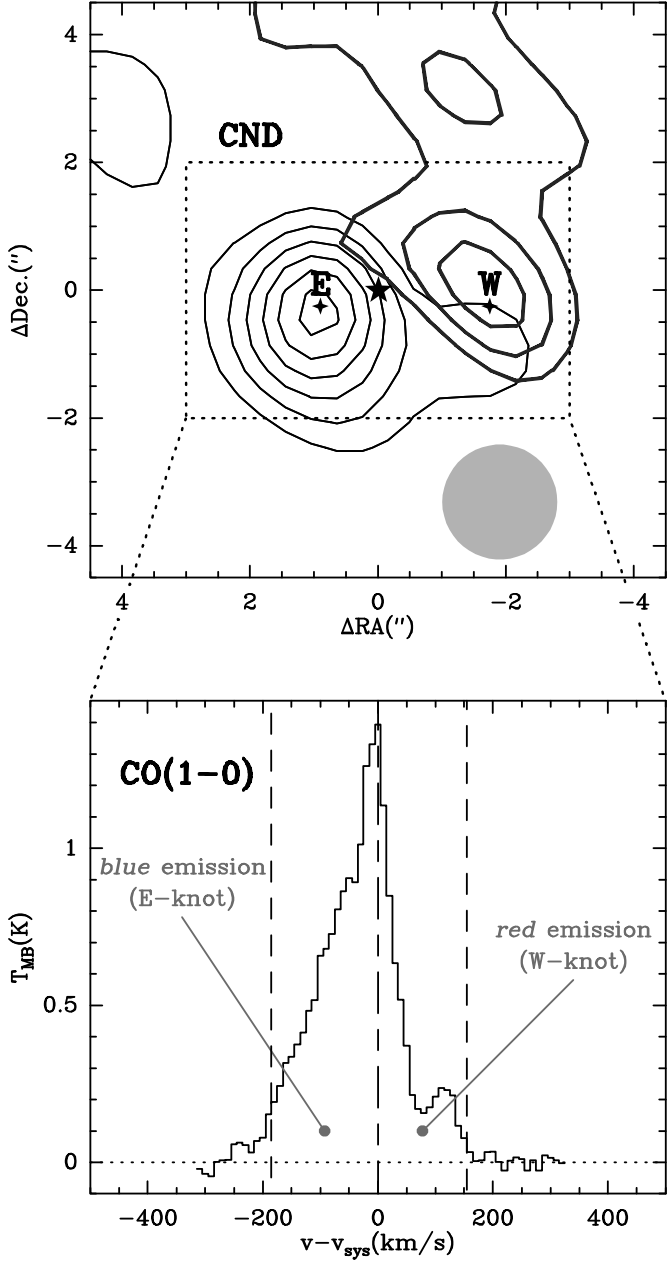} 
\caption{{\sl top panel}: Integrated intensity maps of CO(1--0) toward the 
CND of \object{NGC~1068} obtained for the {\sl blue} (thin contours:from 6$\sigma$ by steps of
3$\sigma$; $\sigma$=0.67~K~km~$s^{-1}$) and {\sl  red} (thick contours: same levels with
$\sigma$=0.47~K~km~s$^{-1}$) emission components as defined in text (see also bottom panel). The
maps have been derived from the data of Schinnerer et al. (\cite{schi00}). The starred marker
highlights the AGN locus. {\sl bottom panel}: Integrated spectrum of CO(1--0) emission  in the CND. The W
and E knots in the CO map correspond, respectively, to the  {\sl red} and {\sl blue} components in
the spectrum.}  
\label{cofig}  
\end{figure} 

\subsection{Morphology of the CND: the Interferometer CO(1--0) Map} 
\label{co-cnd} Fig.~\ref{cofig} represents the CO(1--0) { spatially} integrated spectrum 
of the CND of \object{NGC~1068}. The line emission profile has been obtained 
by integrating the CO(1--0) interferometer data of Schinnerer et al. 
(\cite{schi00}) inside a 6\arcsec$\times$4\arcsec--rectangular region which 
contains the bulk of the CO emission in the CND. According to Schinnerer et 
al. (\cite{schi00})'s estimates, we expect little zero-spacing flux missing 
in the CND integrated spectrum/map. Molecular gas in the CND is not evenly 
distributed around the AGN: two conspicuous knots (denoted as 
E[1\arcsec,0\arcsec] and W[-1.5\arcsec,0\arcsec] knots) form an asymmetrical 
ring around the AGN (See Fig.~1 of Schinnerer et al. \cite{schi00} and Fig.~\ref{PosFig}
in this work). The asymmetrical distribution of molecular gas in the CND is 
{ reflected} by the profile of Fig.~\ref{cofig}: the CO(1--0) 
emission integrated for v$<$v$_{\mathrm{sys}}$ (hereafter, called {\sl blue} 
component) is $\sim$2 times that measured for v$>$v$_{\mathrm{sys}}$ 
(hereafter, called {\sl red} component). As expected for a disk rotating 
around the AGN, the emission coming from the E and W knots roughly 
correspond, respectively, to the {\sl blue} and {\sl red} components defined 
above. This is illustrated in Fig.~\ref{cofig}.

{ The full sizes} of the E and W knots, deconvolved by the 
$1.8\arcsec\times1.8\arcsec$ beam, are alike: FWZP$\simeq$2.2$\arcsec$. 
Therefore we deduce similar areas for the {\sl blue} and {\sl red} emitting 
regions: $\Omega_{\mathrm{source}}\simeq\pi\times~1.1^2$arcsec$^2$=3.8~arcsec$^2$.

\begin{figure*}[!htp]
\includegraphics[width=0.9\textwidth]{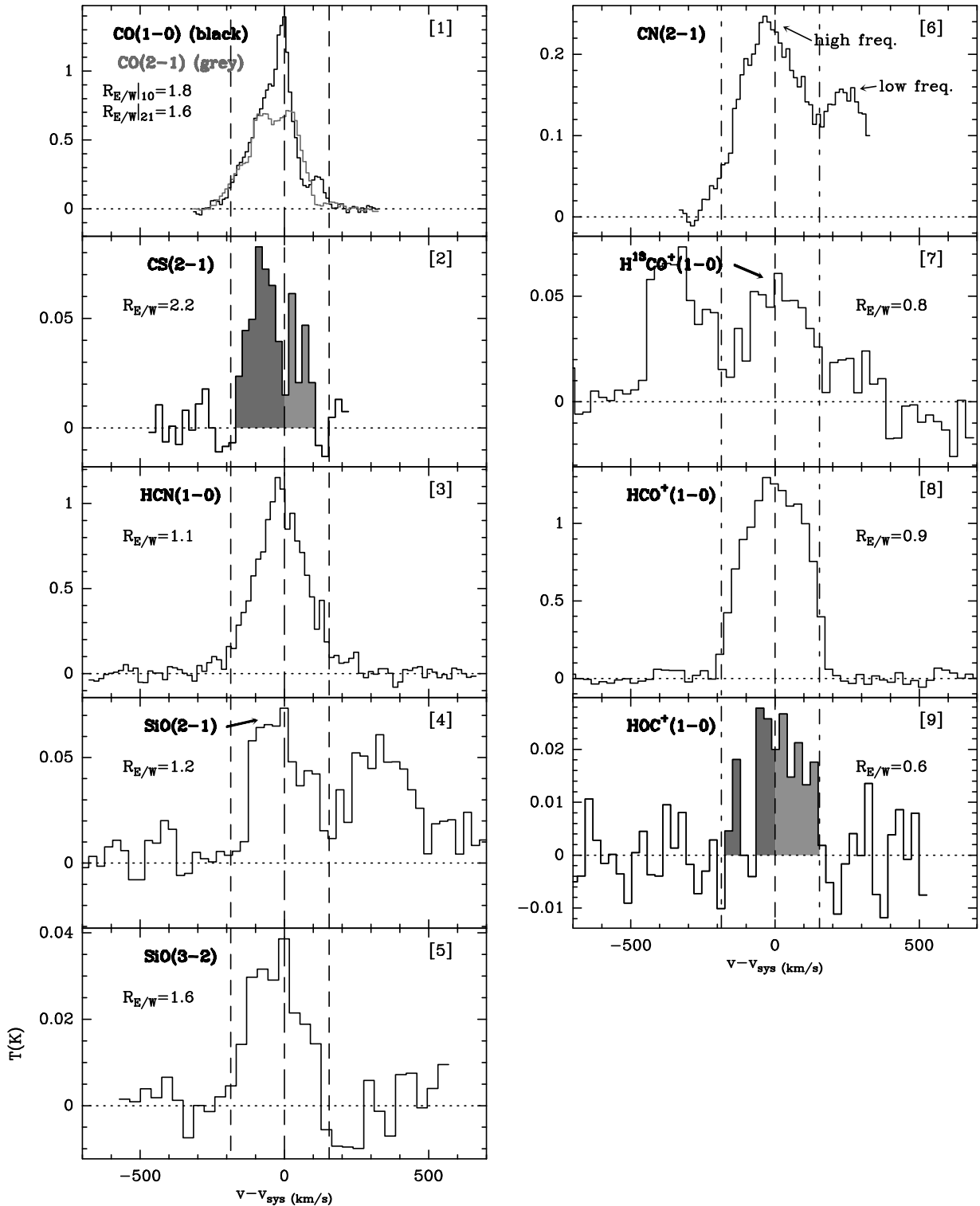} 
\caption{Molecular lines in the CND. Sub-panels are labeled with 
the name of the line displayed. Sub-panels 1 to 3 are derived from 
interferometer data (Tacconi et al. \cite{tacc94}, \cite{tacc97} and 
Schinnerer et al. \cite{schi00}); panels 4 to 9 show single-dish spectra 
observed towards the nucleus (temperatures corrected by dilution effects 
assuming that the emission is coming from the CND). Two vertical point-dashed 
lines at v-v$_{\mathrm{sys}}=$-185~km~s$^{-1}$ and 155~km~s$^{-1}$, delimit 
the {\sl blue} and {\sl red} kinematical components. For each line, the blue-to-red (east-to-west)
average brightness temperature ratio (R$_{\mathrm{E/W}}$) is indicated.}
 \label{CndFigA}  
\end{figure*} 

\begin{figure*}[!htp]
\includegraphics[width=17cm]{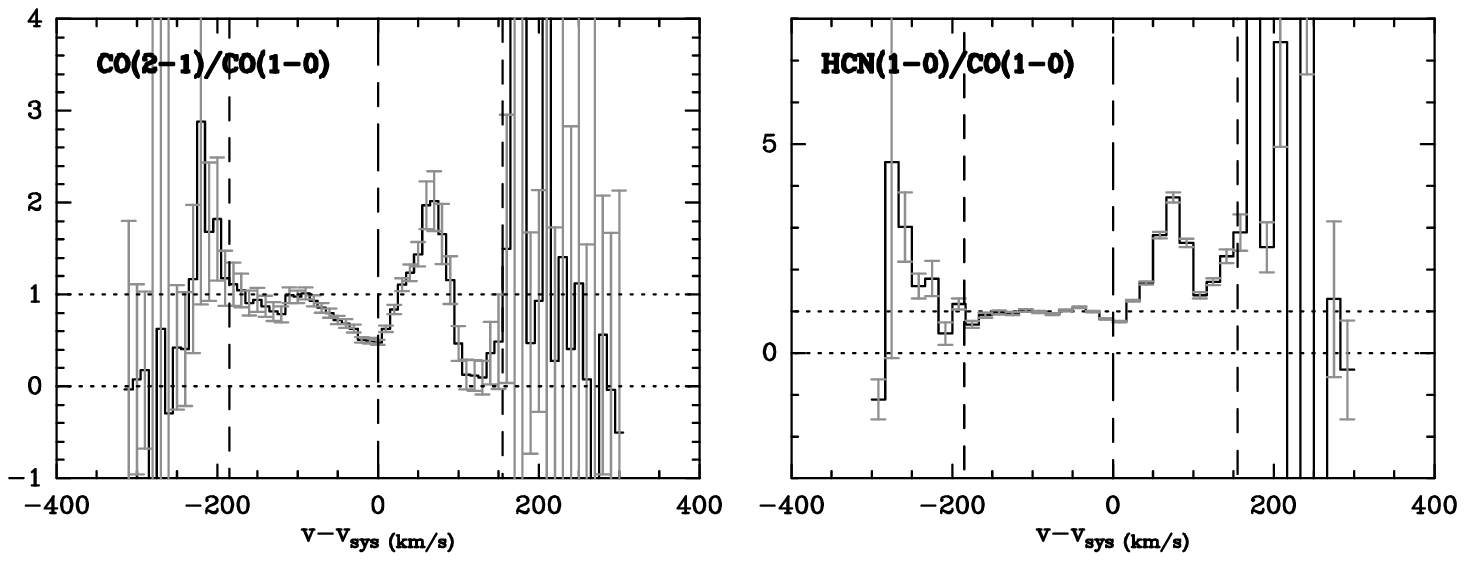} 
 \caption{ Temperature 
ratio profiles derived from spectra of Tab.~\ref{CndFigA}. The left panel 
shows the CO(2--1)--to--CO(1--0) ratio, and the right panel the 
HCN(1--0)--to--CO(1--0) ratio. { Error bars} are $\pm3\sigma$.} \label{CndFigB} 
\end{figure*}

\subsection{Molecular Line Profiles of the CND} \label{profiles} 
Fig.~\ref{CndFigA} displays all the molecular lines observed in this work 
toward the CND of NGC~1068. This includes the 30m spectra of SiO, HOC$^+$, 
HCO$^+$, H$^{13}$CO$^+$ and CN (panels 4--9 in Fig.~\ref{CndFigA}). 
Temperatures have been rescaled assuming that the emission comes from the 
6\arcsec$\times$4\arcsec--rectangular region containing the CND. Furthermore, 
we represent in panels 1--3 of Fig.~\ref{CndFigA}, the CND spectra of CO, CS 
and HCN obtained from published interferometer maps. Similarly to CO (see 
Sect.~\ref{co-cnd}), these CND spectra have been obtained by integrating the 
HCN and CS emission inside the $6\arcsec\times4\arcsec$ rectangular region 
which contains the bulk of the CND flux in both interferometer maps. We can 
redefine more precisely what we call {\sl red} and {\sl blue} velocities, 
ascribed, as argued above, to the W and E knots, respectively: based on the 
observed molecular profiles of Fig.~\ref{CndFigA}, most of the molecular 
emission detected at {\sl red} ({\sl blue}) velocities for all species arises 
within the interval 0$<$v$-$v$_{\mathrm{sys}}<$ 155~km~s$^{-1}$ 
($-$185~km~s$^{-1}<$v$-$v$_{\mathrm{sys}}<0$). { Integrated intensities in the blue and red
components of the spectra are listed in Tab.~\ref{Intensities}.}  

There are noticeable differences between the line shapes of the CND spectra 
shown in Fig.~\ref{CndFigA}. We find line profiles dominated by emission at 
{\sl blue} velocities for CO and CS, while line profiles of HCO$^{+}$, 
H$^{13}$CO$^{+}$ and HOC$^+$ are dominated by {\sl red} emission. As argued 
in Sect.~\ref{sec4}, HOC$^+$ represents an extreme case as the bulk of the 
HOC$^+$ emission is detected at {\sl red} velocities. HCN profiles are rather 
symmetrical with respect to v$_{\mathrm{sys}}$. Finally, the SiO line 
profiles represent a case somewhat intermediate between HCN and CO. These 
differences are quantified in Figure~\ref{CndFigA} { and Tab.~\ref{Intensities}, which show} the
{\sl  blue}--to--{\sl red} ({\sl east}--to--{\sl west}) average brightness temperature 
ratio (R$_{\mathrm{E/W}}$) for all the CND 
spectra { (except for CN(2--1), for which the determination of the blue and red components is
hampered by the partial blending of the lines)}.  
R$_{\mathrm{E/W}}$ { ranges from 2.2$\pm$0.4 
(CS(2--1)) to 0.6$\pm$0.2} (HOC$^+$(1--0)), i.e., { from one extreme to the other}, this 
ratio changes by a significant factor ($\sim$4) among the observed molecules. 
Fig.~\ref{CndFigB} { also illustrates} this result: the HCN(1--0)/CO(1--0) 
temperature ratio is a factor of 2--3 larger for the red component { than  
 for the blue component}. Furthermore, the CO(2--1)/CO(1--0) ratio 
profile, shown in Fig.~\ref{CndFigB}, is also asymmetrical with respect to 
v$_{\mathrm{sys}}$: the (2--1)-to-(1--0) ratio reaches higher-than-one values 
within a 70 kms$^{-1}$ interval at {\sl red} velocities, while it oscillates 
between 0.6 and 0.8 for the {\sl blue} component.

{ Taken together}, these results suggest that there is a chemical differentiation 
between the E and W knots of the CND.

\begin{table}[!htp]
\caption{{ Integrated intensities of the spectra of Fig.~\ref{CndFigA} in the blue (col.2) and
red (col.3) components. Col.4=blue-to-red (east-to-west) ratio of mean temperatures. Errors (in
brackets) are 1-$\sigma$.} }   
\label{Intensities}
{
\input{n1068-usero-tab3.txt}

}
\end{table}
%

\begin{table}[!htp]
\caption{Mean temperatures in the East and West knots of the CND  after correction
for dilution: col. 1
= name of the line; col. 2 = mean temperature in the East-knot;
 col. 3 = idem in the West-knot.
}    
\label{TempLvg}
{
\input{n1068-usero-tab4.txt}

}
\end{table}

\subsection{Molecular Gas Abundances in the CND} \label{SecLvg}

\subsubsection{LVG Models}

We have used single-component Large Velocity Gradient (LVG) models to estimate 
the column densities of the observed molecular species under certain 
assumptions which are the { basis}  of all our calculations. First, we assume 
that the kinetic temperature (T$_{\mathrm{K}}$) of molecular gas in the CND is 50~K. 
This value was derived by Sternberg et al. (\cite{ster94}) from the LVG 
analysis of several CO emission-lines observed toward the CND. Therefore this 
value can be taken as a conservative lower limit for T$_{\mathrm{K}}$. Furthermore, we 
adopt in our calculations an isotopic ratio of $^{12}$C/$^{13}$C=40 (Wannier \cite{wann80}).

   \begin{figure*}[!htp]
\centering
  
\includegraphics[scale=1.154]{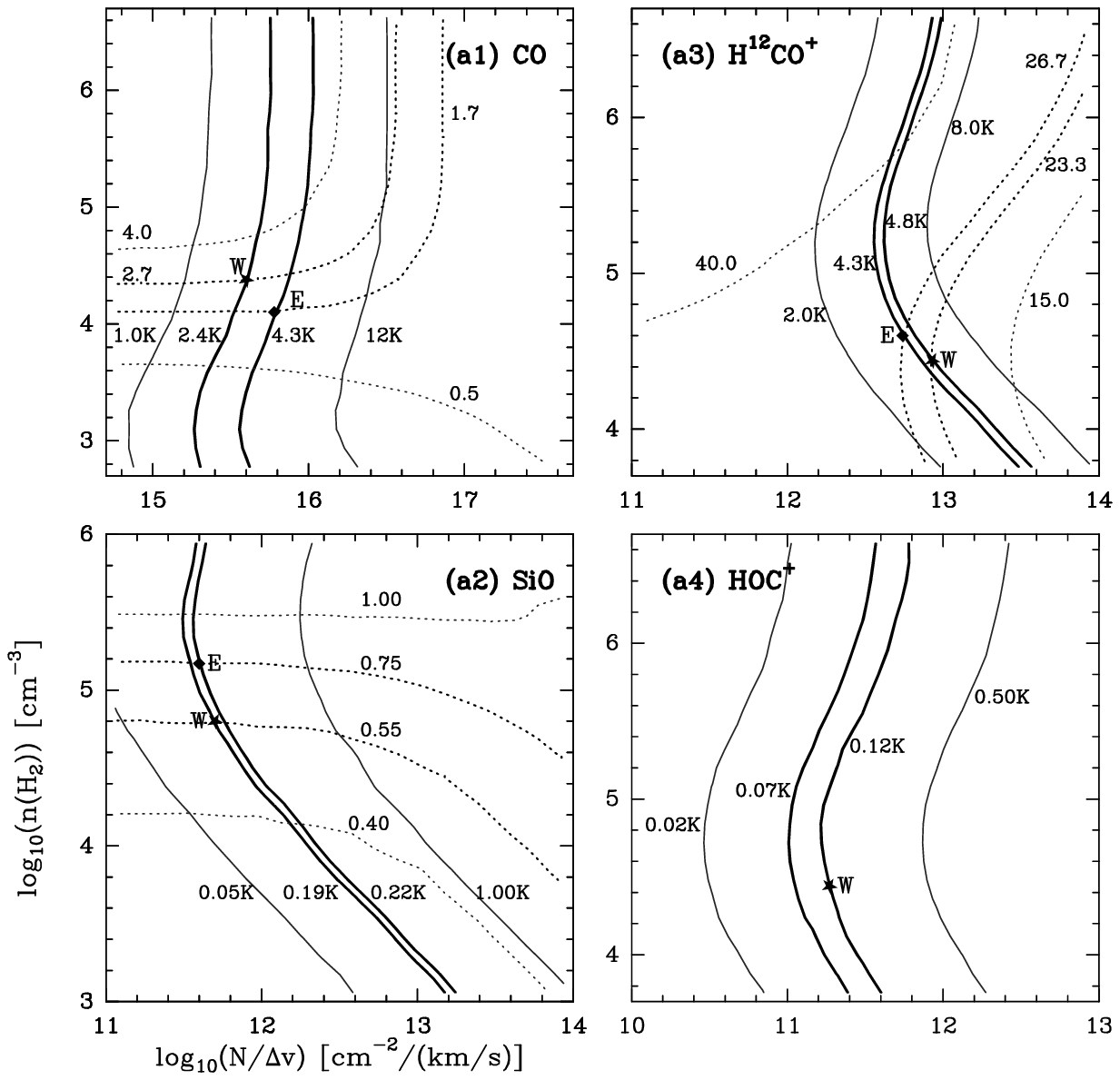}
\caption{LVG estimates for oxygenated species in the E/W knots of the CND. {\sl a1:} for CO,
continuous
 (pointed) curves are contours of constant 1--0 line temperature ((4--3)-to-(1--0) line ratio).
 {\sl a2:} for SiO,  same for 2--1 line temperature ((3--2)-to-(2--1) line ratio).
 {\sl a3:} for H$^{12}$CO$^+$, same for 1--0 line temperature 
([H$^{12}$CO$^+$]-to-H$^{13}$CO$^+$] 1--0 line ratio).
 {\sl a4:} for HOC$^+$, same for 1--0 line temperature.
 Squared (starred) markers show solutions for the East (West) knot.}
\label{LvgFig1}
\end{figure*}


   \begin{figure}[!htp]
    \centering
\resizebox{0.875\hsize}{!}{
\includegraphics[]{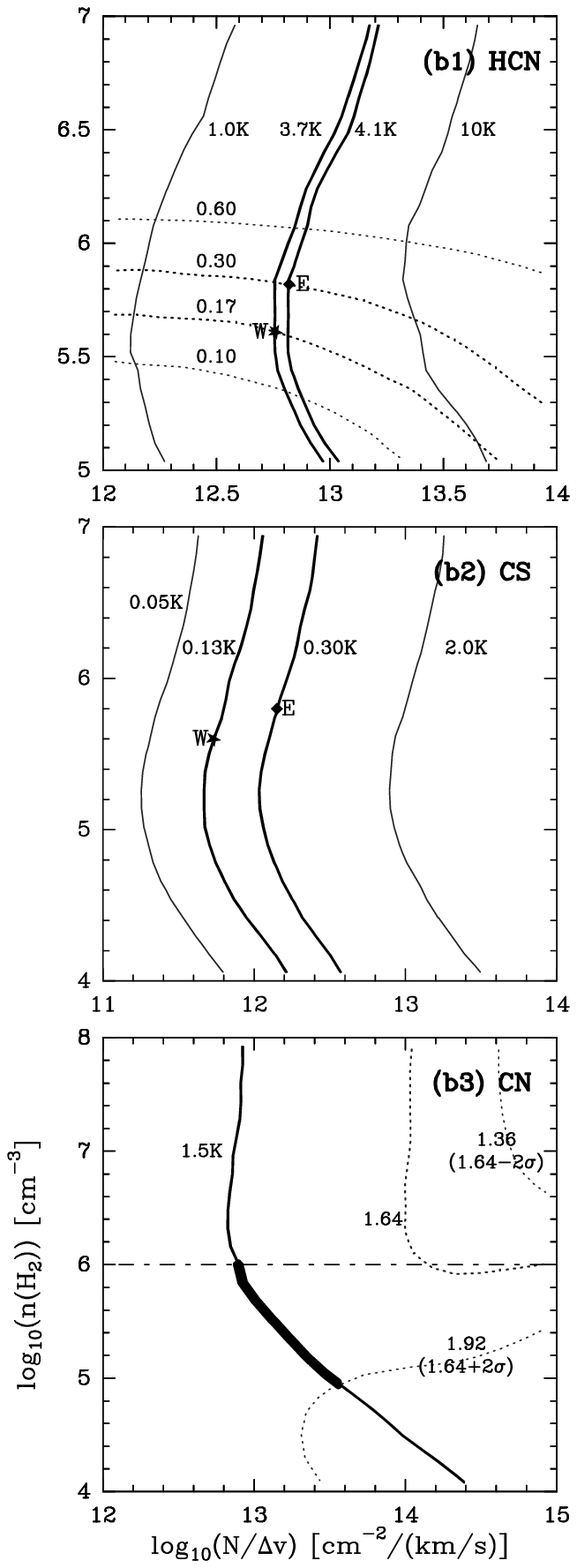}
}
\caption{LVG estimates for non-oxygenated species in the E/W knots of the CND. {\sl b1:} for HCN,
continuous (pointed) lines are contours of constant 1--0 line temperature ((4--3)-to-(1--0) line
ratio). {\sl b2:} for CS, same for 2--1 line temperature. {\sl b3:} for CN, same for 2--1/high
freq. line temperature ((2--1/high freq.)-to-(2--1/low freq.) ratio); { a range of possible
solutions found is highlighted in bold face: we impose n(H$_2$)$<$10$^6$~cm$^{-3}$ for consistency
with the results from others species and allow for a $\pm$2$\sigma$ uncertainty in the measured
ratio}. Markers are as is 
Fig.~\ref{LvgFig1}.}

\label{LvgFig2} 
\end{figure}

It has been previously reported that LVG models of  CO emission in PDR-type environments can lead to
inconsistencies related to spatial fine structure, density and kinetic temperature (see the case of
\object{M~82} in Mao et  al. \cite{mao00}). However, high J-number  transitions (out to
CO(J=7--6)), not available for \object{NGC~1068}, are required to constrain LVG-parameters
sufficiently to search for inconsistencies.  

As argued in Sect.~\ref{co-cnd}, the interferometric CO maps reveal two 
distinct knots (E--W) in the CND. These knots have similar sizes 
($\Omega_{\mathrm{source}}\sim$3.8~arcsec$^2$) and can be identified with two 
adjacent velocity components of emission in the spectra. As discussed in 
Sect.~\ref{profiles}, the relative intensity ratio between these components 
depends on the molecular species. In our calculations we thus give our 
estimates of abundance ratios separately for the E/blue and W/red components. 
All source brightness temperatures (T$_{\mathrm{S}}$(E/W)), listed in 
Tab.~\ref{TempLvg}, have been derived from the CND temperature scale used in 
Fig.~\ref{CndFigA}, corrected by a dilution factor 
$f=\Omega_{\mathrm{CND}}/\Omega_{\mathrm{source}}$.

\begin{table}[!thb]
\caption{LVG results: col. 1= chemical species; col. 2 = Parameters determined from the LVG models
(n: molecular gas densities in cm$^{-3}$; N/$\Delta$v: column densities per
velocity interval in cm$^{-2}$~km$^{-1}$~s; X: chemical abundances relative to  H$_2$;{  we assume
X(CO)=8$\times$10$^{-5}$ and compute the rest of abundances accordingly from column density ratios
relative to CO)}; col. 3= solutions for the East-knot; col. 4 = same for
the West-knot; col 5 = east-to-west ratio of abundances.}
\label{AbundLvg}
{
\input{n1068-usero-tab5.txt}

}
\end{table}

The range of LVG solutions (n(H$_2$), N/$\Delta$v) are determined 
straightforwardly for SiO, CO, HCN and CN from the observed line ratios and 
the source brightness temperatures. In the case of SiO, we fit the 
(3--2)-to-(2--1) ratio and the 2--1 line source temperature. Correction for contamination from the
ring is taken into account for SiO(2--1) (also for HCO$^+$(1--0); see below). For CO and HCN 
we use the (4--3)-to-(1--0) line ratios and the 1--0 line source 
temperatures; 4--3 line temperatures of CO and HCN are derived from 
single-dish data published by Tacconi et al. (\cite{tacc94}). In the case of 
CN, we fit the ratio of the two fine structure lines and the low-frequency 
line source temperature. However, and due to partial blending of the two fine 
groups, LVG solutions refer to global abundances with no distinction between 
red and blue velocity components.
   The LVG solution for H$^{12}$CO$^+$ is 
obtained by fitting both the H$^{12}$CO$^+$-to-H$^{13}$CO$^+$ temperature 
ratio measured for the 1--0 line and the H$^{12}$CO$^+$(1--0) source 
temperature. We have implicitly assumed that the derived density solution 
n(H$_2$) can be considered as common for both H$^{12}$CO$^+$ and 
H$^{13}$CO$^+$.
In the case of HOC$^+$, LVG estimates are only obtained in the W knot,
since the signal-to-noise ratio of the integrated emission at blue velocities
(E knot) is too low ($<$5).  
 The estimate of N/$\Delta$v for 
HOC$^+$, only observed in the 1--0 line, { rests on} the assumption of a 
value for n(H$_2$), here taken from H$^{12}$CO$^+$. This approach is 
justified as HOC$^+$ and HCO$^+$ are known to be 
formed/destroyed in chemical reactions taking place in the same gas clouds.    
Similarly, N/$\Delta$v values for CS are derived assuming for this molecule 
the same gas density inferred from HCN in order to fit the CS(2--1) source 
brightness temperature.

Figures \ref{LvgFig1} and \ref{LvgFig2}, and Tab.~\ref{AbundLvg} summarize the 
results of LVG calculations for CO, HCN, CS, CN, SiO, HCO$^+$ and HOC$^+$. 
Normalized with respect to N(CO), the column densities of SiO, HCO$^+$ and 
HOC$^+$ (i.e., N(SiO)/N(CO), N(HCO$^+$)/N(CO) and N(HOC$^+$)/N(CO)) are 
$\geq$2--3 larger in the W knot than in the E knot. In contrast, N(HCN)/N(CO) 
and N(CS)/N(CO) column density ratios are similar in the two knots within a 
25\% uncertainty.  These abundance ratios are 
reflecting the asymmetries of the spectra discussed in Sect.~\ref{profiles}, 
suggestive of an uneven processing of molecular gas in the CND.

 As a byproduct of LVG models for CO, we have estimated the X$\equiv$N(H$_2$)/I(CO) conversion
factor for the molecular gas in the CND. Assuming a range of abundance ratios
[CO]/[H$_2$]$\sim$5$\times$10$^{-5}$--10$^{-4}$, we infer a X value of
3-6$\times$10$^{19}$~cm$^{-2}$/(K~km~s$^{-1}$), i.e.,  $\sim$4--8 times smaller than the canonical
value X=2.2$\times$10$^{20}$ (Solomon \& Barrett \cite{solo91}). Unless CO is underabundant by a
similar factor (a scenario invoked by Sternberg et al. \cite{ster94} in the oxygen depletion models
clearly invalidated by the results of our work; see section~\ref{sec6}) we conclude that the X
conversion factor is lower in the CND of \object{NGC~1068}. Similar deviations have been previously
reported in other galactic central regions  (Dahmen et al. \cite{dahm98} and references therein).
 This might reflect the failure of some of the basic hypothesis
that support the canonical value. In particular, the strong gravitational forces near galactic
nuclei may prevent molecular clouds from reaching virialization.

\section{Chemistry of Molecular Gas in the CND of \object{NGC~1068}} \label{sec6}

To give further insight into the chemistry of molecular gas in the CND we have 
compared, for a common set of abundance ratios, the values measured in \object{NGC~1068} 
with { those} observed in a {\sl reference} galactic region. Here we take as 
`zero-point' environment the ``Extended Ridge'' of OMC-1 (OER) (Blake et al. 
\cite{blak87}), a relatively quiescent molecular region whose chemistry has 
been described as intermediate between the { one} typical of cold dark clouds and 
that of warm cores (Sutton et al. \cite{sutt95}). The choice of the OER as a 
{\sl reference} region is also motivated by the similarity of physical 
parameters of molecular gas density (n(H$_2$)$\sim$10$^{4}$--10$^{5}$~cm$^{-3}$) and 
kinetic temperature (T$_\mathrm{K}$$\sim$50~K) in the CND and in the OER. Therefore, 
significant differences in the abundance ratios of `critical' tracers between 
the CND and the OER can be mostly attributed to different chemistries being 
at work in these regions. We will also use the OER as the `zero point' basis 
to extrapolate the abundance ratios in the case of oxygen depletion models { (Ruffle et
al. \cite{ruff98}).}

\begin{table*}[!hbt]
\caption{Abundance ratios predicted/observed in different molecular regions: the E-knot of the CND
of \object{NGC~1068}, same for the West-knot; a prototypical XDR { (Lepp \& Dalgarno \cite{lepp96};
Yan \& Dalgarno \cite{yan97})}, the Orion Extended Region (OER) { (Blake et
al. \cite{blak87})}, and the OER
corrected with oxygen depletion { (Ruffle et al. \cite{ruff98}).} }
\label{chemods}
\input{n1068-usero-tab6.txt}
\end{table*}

We list in Table~\ref{chemods} the following set of abundance ratios: N(HCN)/N(CO), N(CS)/N(CO),
N(HCN)/N(HCO$^+$), N(CN)/N(HCN), N(SiO)/N(CO) and N(HCO$^+$)/N(HOC$^+$). These abundance ratios can be 
significantly different depending on the chemical environment. As argued 
below, an evaluation of these ratios { allows} us to compare the chemical status 
of the CND and the OER with the predictions of models invoking either 
oxygen-depletion or X-ray driven chemistry:

\begin{itemize}

\item

Our observations provide new constraints for oxygen-depletion models first 
proposed by Sternberg et al.~(\cite{ster94}) as an explanation for the high 
HCN/CO ratio measured in the CND of \object{NGC~1068}. This scenario is supported by 
X-ray and ultraviolet observations of the hot-ionized gas in the narrow-line 
region of \object{NGC~1068} (Marshall et al. \cite{mars93}, Ogle et al. \cite{ogle03}). 
With the inclusion of dust-grain chemistry in time-dependent models, 
Shalabiea \& Greenberg (\cite{shal96}) were able to fit at `early times' 
(t$\simeq$10$^6$~yr) HCN/CO$\sim$a few 10$^{-3}$ with values less restrictive 
for the oxygen depletion. The overall consequences of selective 
oxygen-depletion in the chemistry of molecular clouds have been more extensively 
studied in the framework of gas-phase (Terzieva et al. \cite{terz98}; Ruffle 
et al. \cite{ruff98}) and gas-grain chemical models (Shalabiea 
\cite{shal01}). The primary effect of an oxygen underabundance is a reduced 
formation of CO. The fraction of carbon not consumed in the CO synthesis is 
then increased and it can thus enhance the abundances of some carbonated 
species, such as HCN, CS or CN; on the contrary, abundances of oxygen-bearing 
{ species} are expected to be lower. This decrease is less important for HCO$^+$, as { in this
case} a 
lower oxygen abundance { is  mostly} balanced by the increase of available 
carbon.

As shown in Table~\ref{chemods}, the measured HCN/CO ratio in the CND of \object{NGC~1068} 
($\sim$a few 10$^{-3}$) is 1 order of magnitude larger than that derived for 
the OER.	
 Oxygen depletion models can fit the HCN-to-CO ratio of the CND with an oxygen 
depletion of [O]$_{\mathrm{CND}}$/[O]$_{\mathrm{OER}}\sim$1/2. However, this value of oxygen
depletion would
lead to 
large HCN/HCO$^+$ ratios ($\sim$25) which are at odds with the low ratios 
($\sim$1) of the CND. Furthermore, these models predict a significant 
enhancement of CN due to the reduction of O~\textsc{i} which is an important source of 
CN destruction (Bachiller et al. \cite{bach97}). Here also the CN-to-HCN 
ratio in the oxygen depletion models solution ($\sim$20) is nearly one order 
of magnitude larger than the CND ratio ($\sim$1--5, i.e., slightly above the 
OER standards). Finally, the predicted variation for the CS/CO ratio is 
marginal ($\times$1.4) in the adopted oxygen-depletion solution, leading to 
values similar to that reported for the CND: CS/CO$\sim$2$\times$10$^{-4}$.

\item

Lepp \& Dalgarno (\cite{lepp96}) proposed an alternative explanation of the 
high HCN/CO ratio measured in the CND of \object{NGC~1068}: X-rays coming from the 
central engine may significantly enhance the abundance of HCN in the 
neighbouring molecular gas. Thus, the HCN/CO ratio measured in \object{NGC~1068} can be 
easily accounted for. In a XDR chemistry some diatomic species, such as CN 
and OH are particularly robust (Lepp \& Dalgarno \cite{lepp96}). Moreover, large 
abundances of OH favour the formation of CO$^+$ and H$_2$O  
(Sternberg et al. \cite{ster96}); these species 
take part directly in the production of large quantities of HCO$^+$. The 
abundances of HCN, CN and HCO$^+$ { simultaneously reach} their peak values at 
similar depths inside XDR (Yan \& Dalgarno \cite{yan97}). The XDR model of 
Yan \& Dalgarno (\cite{yan97}) predicts an average CS/CO abundance ratio of
 1--5$\sim$10$^{-4}$   
for the range of depths inside the XDR that are expected to dominate the 
emission of molecular gas. As summarized in Table~~\ref{chemods}, the HCN/CO, 
HCN/HCO$^{+}$, CN/HCN and CS/CO abundance ratios predicted by XDR models { (see Lepp \& Dalgarno
\cite{lepp96} for the three first ratios; the CS/CO ratio has been estimated from Yan \& Dalgarno
\cite{yan97}) }
are 
in close agreement with the corresponding values estimated for the CND of 
\object{NGC~1068}.

\end{itemize}

In summary, while oxygen depletion models are able to fit the HCN/CO ratio 
measured in the CND of \object{NGC~1068}, the adopted solution leads to 
HCN/HCO$^{+}$ and CN/HCN abundance ratios which are excessively large 
compared to that actually measured for the CND. In contrast, the models 
invoking XDR chemistry explain naturally the ratios measured in \object{NGC~1068}; these 
values depart significantly from the standard reference values of the OER. In 
the following sections we discuss how the detection of high abundances of SiO 
and HCO$^{+}$ in the CND of \object{NGC~1068} add supporting evidence to the XDR 
chemistry scenario.

\subsection{SiO in XDR} \label{SiOXdr}

{ As is shown} in Tab.~\ref{chemods}, the SiO-to-CO abundance ratio measured 
toward the CND of \object{NGC~1068} is high by normal galactic standards: 
N(SiO)/N(CO)$\sim$6$\times$10$^{-5}$-1.2$\times$10$^{-4}$. The normalized SiO 
column densities toward the CND are at least one order of magntitude larger 
than the upper limit derived for the OER ($<$7$\times$10$^{-6}$). Assuming an 
absolute abundance for CO of X(CO)=8$\times$10$^{-5}$, this implies 
X(SiO)=5$\times$10$^{-9}$-1.0$\times$10$^{-8}$. As discussed in 
Sect.~\ref{sec3.1}, a significant enhancement of SiO in molecular gas has 
been attributed to heavy shock processing of grains in starburst galaxies 
where values of X(SiO) up to a few 10$^{-10}$ have been reported on scales of 
several hundred pc (Garc\'{i}a-Burillo et al. \cite{buri00, buri01}). The CND 
abundances of SiO estimated here are significantly larger than { those} reported 
for starbursts, however; this is further evidence that silicon chemistry in 
the CND is not being driven by star formation. In contrast, the estimated SiO 
abundances in the starburst ring of \object{NGC~1068} 
(X(SiO)$\sim$2-3$\times$10$^{-10}$) are in close agreement with SiO 
abundances measured in starbursts on similar spatial scales.

Alternatively, it has been suggested that X-ray irradiated dust grains can 
enhance silicon chemistry in gas phase. X-rays are able to heat very small 
silicate grains (10~\AA), subsequently leading to their evaporation and to an 
enlargement of the Si gas-phase fraction (Voit \cite{voit91}). Most 
remarkably, the nucleus of \object{NGC~1068} shows a strong Fe~K$\alpha$ line (Ogle et 
al. \cite{ogle03} and references therein). The bulk of the 6.4~keV line of 
Fe~\textsc{i} most likely comes from fluorescence in the Compton-thick molecular gas 
torus of \object{NGC~1068}. The detection of strong Fe~K$\alpha$ line emission is 
therefore an indication that large column densities of molecular gas are 
being processed by X-rays. In a precedent study, Mart\'{\i}n-Pintado et al. 
(\cite{mart00}) found a correlation between the intensity of the Fe 6.4~keV 
line and the derived abundance of SiO in the \object{Sgr~A} and \object{Sgr~B}
 molecular 
complexes at the Galactic Center.

\subsection{HOC$^+$ in XDR} \label{HocXdr}

According to the estimates of Sect.~\ref{SecLvg}, HOC$^+$ abundances derived 
for the CND of \object{NGC~1068} are the largest ever measured in interstellar medium: 
X(HCO$^+$)/X(HOC$^+$)$\sim$30--80. These low ratios are
in direct contrast with { those} typically measured in galactic dense molecular 
clouds where values from $\sim$6000 to $\sim$300 have been reported thus far (Apponi et al. 
\cite{appo97}, \cite{appo99}). Most interestingly, the lowest value found by 
Apponi et al. (\cite{appo99}) corresponds to the Orion bar, a prototypical PDR. 
Very low ratios ($\sim$50-120) have been recently found in the prototypical 
PDR \object{NGC~7023} (Fuente et al. \cite{fuen03}). As argued below, we 
propose that low  R$\equiv$X(HCO$^+$)/X(HOC$^+$) ratios can be explained for 
molecular clouds with high ionization degrees, either in XDR or in PDR.

   \begin{figure}[!htp]
\resizebox{\hsize}{!}
{
\includegraphics[angle=0]{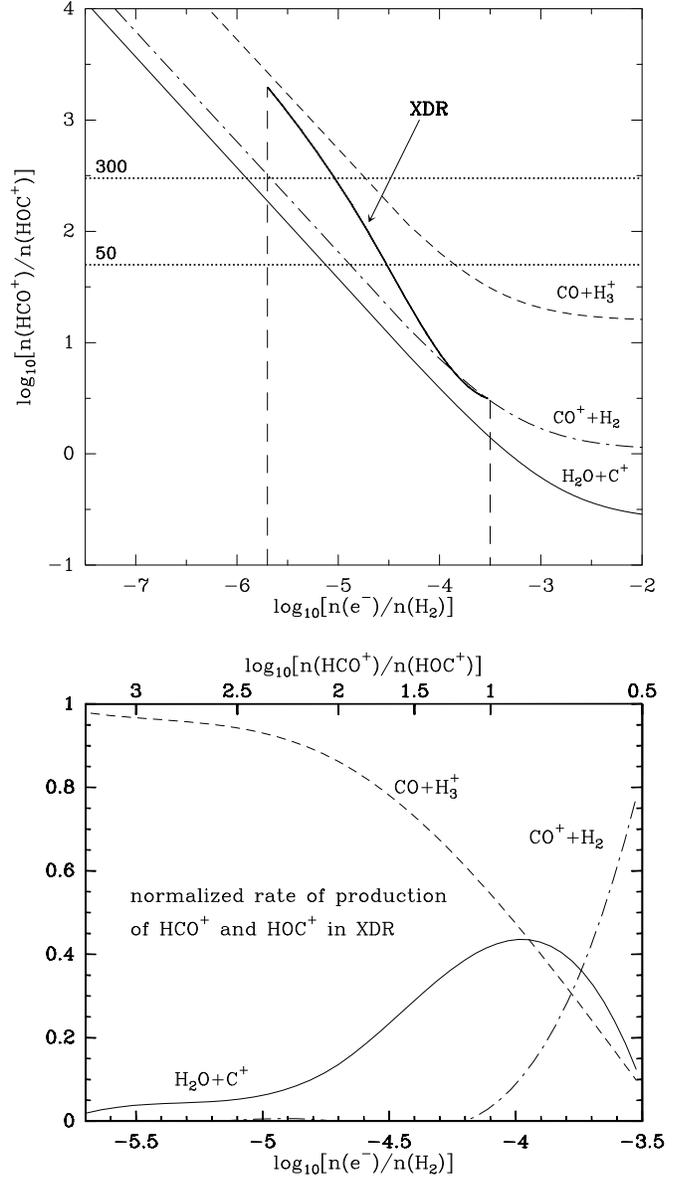}
}
\caption{{\sl Top panel}: steady state HCO$^+$-to-HOC$^+$ abundance ratio as a function of the
ionization degree of molecular gas. Curves for single formation paths are plotted; the thick line
shows the predicted ratio for a XDR chemistry. {\sl Bottom panel}: fraction of  HCO$^+$ and HOC$^+$
molecules formed along each chemical path in a XDR
chemistry.}           
\label{HocFig} 
\end{figure}

The fast hydrogen-catalyzed isomerization of HOC$^+$ into HCO$^+$ usually 
shifts the equilibrium between both species towards significantly lower 
abundances of HOC$^+$. However, as suggested by Smith et al. (\cite{smit02}), 
the isomerization process converting HOC$^+$ into HCO$^+$ could be 
compensated by the destruction of HCO$^+$ due to interaction with electrons. 
This process is likely to be enhanced at high electron densities 
(X(e$^-$)$\sim10^{-5}$). The latter could explain why { the} lowest R values have been 
measured in galactic PDR (Apponi et al. \cite{appo99}; Fuente et al. 
\cite{fuen03}). Furthermore, the X(HCO$^+$)/X(HOC$^+$) ratio at equilibrium is 
also sensitive to the dominant mechanism of HCO$^+$/HOC$^+$ formation: the 
more efficient is the relative production of HOC$^+$, the lower is the 
ionization degree required to reach a certain R ratio. Typical paths for the 
formation of HCO$^+$/HOC$^+$ are { (Apponi et al. \cite{appo97} and references therein)}:
\begin{eqnarray}
\mathrm{H}^+_3+\mathrm{CO}&\longrightarrow&\mathrm{HOC}^+/\mathrm{HCO}^++\mathrm{H}_2
\label{Eqnh3+}\\ 
\mathrm{CO}^++\mathrm{H}_2&\longrightarrow&\mathrm{HOC}^+/\mathrm{HCO}^++\mathrm{H}
\label{Eqnco+}\\
\mathrm{H_2O}+\mathrm{C}^+&\longrightarrow&\mathrm{HOC}^+/\mathrm{HCO}^++\mathrm{H}
\label{Eqnh2o}
\end{eqnarray}

 The branching ratio for the net production of nascent HOC$^+$, hereafter 
denoted by $\alpha$,
depends on the particular formation pathway. The value of $\alpha$ is 0.06  
for reaction
\ref{Eqnh3+}, 0.48 for \ref{Eqnco+} and 0.8 for \ref{Eqnh2o}. In a real case 
scenario the three reactions { coexist}, and thus the equivalent branching 
ratio, $\alpha_{\mathrm{eff}}$, is an average of the individual $\alpha$, 
weighted by the fraction of HCO$^+$ and HOC$^+$ particles that are actually 
formed following a certain pathway. { We have derived how R depends on 
the ionization degree of molecular gas, separately, for the different 
reactions, assuming the rate coefficients given by Smith et al. (\cite{smit02}) (Fig.~\ref{HocFig},
top panel)}.  We find that in order to obtain values of
R$\sim$50-300 high  
ionization degrees are needed: X(e$^-$)$\sim$10$^{-6}$-10$^{-4}$. These high 
electronic abundances are typically reached in XDR (Lepp \& Dalgarno \cite{lepp96}, 
Maloney et al. \cite{mall96}). Fig.~\ref{HocFig}
illustrates also the relation between the dominant formation path and the isomer 
ratio: less extreme ionization degrees are needed to reach low R ratios if 
the predominant reaction has a large $\alpha$. On the other hand, much higher 
ratios $\sim$300-6000, like that { typically} measured in molecular clouds, can be 
easily accounted for if electronic abundances approach standard levels 
$<$10$^{-6}$--10$^{-7}$.

 We have also derived the dependence of $\alpha_{\mathrm{eff}}$ { on} the ionization degree for an
adopted XDR model { (see the curve for R in the top panel of
Fig.~\ref{HocFig})}. 
 The abundances of all molecular species, { contributing to} (\ref{Eqnh3+}), 
(\ref{Eqnco+}) and (\ref{Eqnh2o}),  have been taken from Maloney et al. 
(\cite{mall96}), except for { CO$^{+}$, whose} abundance curve is taken from the 
PDR model of Sternberg et al. (\cite{ster95}). Values of R$\sim$30--80, like that 
measured in the CND of \object{NGC~1068}, can be easily accounted for assuming an 
average ionization degree of X(e$^-$)$\sim$10$^{-5}$ for the { bulk of molecular 
gas.

 The relative weight } of the 3 formation paths of HOC$^+$ in a typical XDR 
is also represented, as a function of X(e$^-$), in Fig.~\ref{HocFig} { (bottom panel)}. While 
reaction (\ref{Eqnh3+}) clearly dominates the balance for 
X(e$^-$)$<$10$^{-5}$, reactions involving H$_2$O (\ref{Eqnh2o}) and 
CO$^{+}$(\ref{Eqnco+}) are predominant for X(e$^-$)$>$10$^{-4}$. Recent 
observations of galactic PDR (Fuente et al. \cite{fuen03}; Rizzo et al. 
\cite{rizz03}) have confirmed that low HCO$^+$/HOC$^+$ ratios are indeed 
correlated with large abundances of CO$^+$ and/or H$_2$O.

\subsection{Anisotropic X-ray Illumination of the CND?} \label{anisotropic}

   \begin{figure*}[!tb]
   \sidecaption
    \centering
\resizebox{12cm}{!}
{
\includegraphics[angle=0]{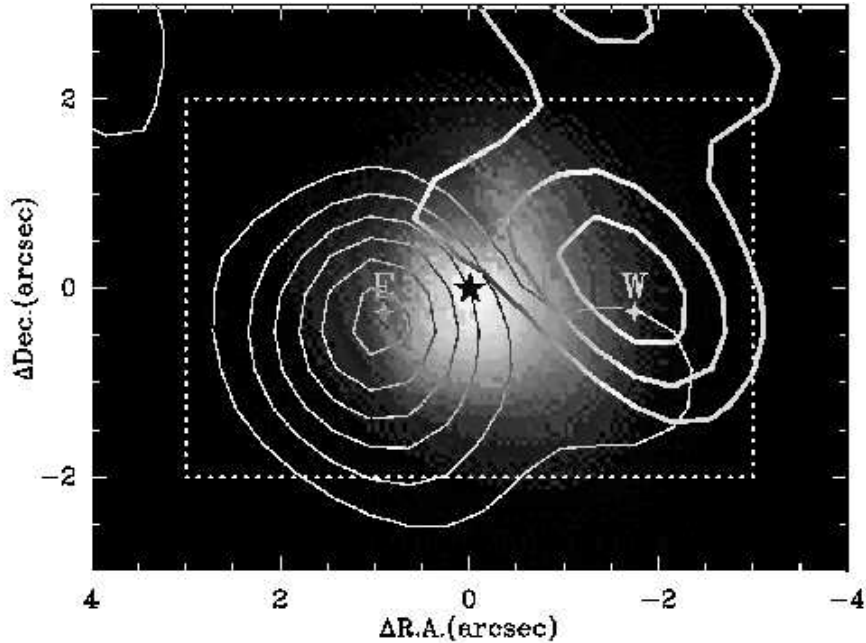}
}
\caption{X-ray emission and molecular gas in the CND: overlay of the distribution of hard X-ray
emission in the 6-8~keV band (gray scale adapted from Ogle et al. \cite{ogle03}: whiter shades
stand for stronger emission) and the CO(1--0) integrated emission as in Fig.~\ref{cofig}. The AGN
locus is highlighted by the starred marker.}             
\label{FeiFig} 
\end{figure*}

The results of this work strongly favour an overall scenario where the CND of 
\object{NGC~1068} has become a giant XDR. It is tempting to speculate if X-ray driven 
chemistry can also { explain the mild} but systematic 
differences in the molecular abundances of SiO, HCO$^{+}$, HOC$^{+}$ and HCN 
between the E and the W knots of the CND. Figure~\ref{FeiFig} shows the 
distribution of X-ray emission inside the 6-8~keV band in the CND of \object{NGC~1068} 
(adapted from Ogle et al.~\cite{ogle03}). X-ray emission in this energy band 
is dominated by the Fe~\textsc{i} K$\alpha$ line at 6.4~keV in \object{NGC~1068}. There 
is a 10$\arcsec$ extended emission (not shown in Figure~\ref{FeiFig}) which 
corresponds to the ionization cones. The strongest component, however (shown 
in  Figure~\ref{FeiFig}), { should} be tracing the illuminated inner wall of the 
CND torus. Most interestingly, this figure shows tantalizing evidence of a different degree of 
penetration of X-rays into the E/W knots: the western side of 
the molecular torus, corresponding to the inner wall of W knot, seems to be 
more illuminated than its eastern counterpart. This would be in agreement 
with the reported chemical differentiation seen between the E/W molecular 
knots.  A difference in the attenuating 
column densities, estimated from CO, exists between the two CND knots: on 
100~pc scales, N(H$_2$)$|_{\mathrm{East}}$/N(H$_2$)$|_{\mathrm{West}}\sim$2. 
However, we do not know if these or
even greater differences hold at smaller scales which are probably more 
relevant to probe X-ray absorption by neutral gas in the torus. In this 
context, it is however suggestive to note that the strongest H$_2$O megamasers, which 
are collisionally excited in the warmest region of the CND illuminated by X 
rays (Neufeld et al. \cite{neuf94}), are mostly located in the western side of the molecular torus
(Greenhill \& Gwinn \cite{gree97}).  

High-resolution interferometer observations will give a sharp view of 
molecular abundance changes inside the CND at small scales for {\sl critical} 
tracers such as SiO, CN and HOC$^+$.  A detailed comparison of these maps with 
the Chandra images of the CND may help to constrain this scenario.


 \section{Conclusions}
\label{sec7}

We summarize the main results obtained in this work as follows:

\begin{itemize}

\item

We report on the detection of significant SiO(3--2) and SiO(2--1) emission in 
the 200~pc circumnuclear disk of \object{NGC~1068}. The large overall 
abundance of SiO in the CND ($\sim$(5-10)$\times$10$^{9}$) cannot be 
explained by shocks driven by star formation on molecular gas as there is 
counter-evidence of a recent starburst in the nucleus of \object{NGC~1068}. 
While SiO emission is also detected over the starburst ring, we 
estimate that SiO abundances there are 10 times lower than { those} measured in the CND. These lower 
abundances of SiO are in close agreement with that measured in starbursts on similar spatial scales,
however.

\item

 We also report on the first extragalactic detection of the reactive ion 
HOC$^+$. Most remarkably, the estimated HCO$^+$/HOC$^+$ abundance ratio in 
the nucleus of \object{NGC~1068}, $\sim$30--80, is the
smallest ever measured in molecular gas. The line profile of HOC$^+$ is 
markedly asymmetrical with respect to v$_{\mathrm{sys}}$: HOC$^+$ emission is 
mostly detected at {\sl red} velocities. Whatever process is responsible for 
the enhancement of this reactive ion, it seems to be unevenly efficient 
inside the CND.

\item

Results from additional mm-observations have served for estimating
abundances of CN, HCO$^+$, HOC$^+$, H$^{13}$CO$^+$ and 
HCO. These estimates are complemented by a 
re-evaluation of molecular abundances for HCN, CS and CO, based on previously 
published single-dish and interferometer observations of \object{NGC~1068}. 
While models invoking oxygen depletion in molecular gas successfully fit the 
HCN/CO ratio measured in the CND, they fail to account for our estimates of 
the HCN/HCO$^{+}$ and CN/HCN abundance ratios. On the contrary, XDR models 
can  { simultaneously explain} these ratios. The
detection of high abundances of SiO and HOC$^{+}$ in the CND of \object{NGC~1068} gives 
further support to the XDR chemistry scenario. The processing of 10~$\AA$ dust grains 
by X-rays, as a mechanism to enhance silicon chemistry in gas phase, would 
explain the large SiO abundances of the CND. Finally, we have shown that the 
low  HCO$^+$/HOC$^+$ ratios measured in the CND can be explained if molecular 
clouds have the high ionization degrees typical of XDR (X(e$^-$)$\sim 
10^{-6}$-$10^{-4}$). An examination of the different formation paths of 
HOC$^{+}$ suggests that reactions involving H$_2$O and/or CO$^{+}$ would be 
the predominant precursors of HOC$^{+}$ in XDR.

\item

The XDR scenario could also provide an explanation for the different abundances of SiO, HCO$^+$ and,
especially, of HOC$^+$ measured in the E and W knots. The Chandra images of the CND in the 6-8~keV
band, dominated by the emission of the Fe \textsc{i} K$\alpha$ line, show tantalizing evidence of a different degree
of penetration of hard X-rays into the E and W knots. This suggests that larger columns of
molecular gas are being processed by X-rays in the W knot.

\end{itemize}


\begin{acknowledgements}
We acknowledge the IRAM staff from  Pico Veleta and Granada for help provided during the
observations.
We wish to { thank A. Rodr\'{\i}guez-Franco for his} support during the observations. We also
wish to thank  E. Schinnerer and L. J. Tacconi for providing their interferometer data.  This
research has made use of NASA's Astrophysics Data System (ADS) and the NASA/IPAC Extragalactic
Database (NED).  This paper has been partially funded by the Spanish MCyT under projects
DGES/AYA2000-0927,  ESP2001-4519-PE, ESP2002-01693 , PB1998-0684, ESP2002-01627 and AYA2002-10113E.
\end{acknowledgements}

\end{document}

%% file: n1068-usero-tab1.txt
\begin{tabular}{lr@{.}lcccr@{/}l}
\hline
\hline
\noalign{\smallskip}
\multicolumn{8}{c}{\bf New Observations}\\
\noalign{\smallskip}
\hline
\noalign{\smallskip}
Line	&Freq&(GHz) & Obs. dates &Beam (\arcsec) &$\eta_{\mathrm{B}}$&
T$_{\mathrm{rec}}$&T$_{\mathrm{sys}}$ (K)\\ 
\noalign{\smallskip}
\hline
\noalign{\smallskip}

H$^{12}$CO(3/2-1/2,2-1)	& 86&670 & Jun00/Aug02&28 &0.82 &70&130\\
H$^{13}$CO$^+$(1--0)	& 86&754 & Jun00/Aug02&28 &0.82 &70&130\\
SiO(2--1)		& 86&847 & Jun00/Aug02&28 &0.82 &70&130\\
H$^{12}$CO$^+$(1--0)	& 89&189 & Jan01/May01&27 &0.81 &60&120\\
HO$^{12}$C$^+$(1--0)	& 89&487 & Jan01/May01&27 &0.81 &60&120\\
SiO(3--2)		& 130&269& Jun00/Aug02&19 &0.77 &125&225\\
CN(2--1)		& 226&875& Aug02      &11 &0.58 &120&390\\
\noalign{\smallskip}
\hline
\noalign{\bigskip}
\noalign{\bigskip}
\hline
\hline
\noalign{\smallskip}
\multicolumn{8}{c}{\bf Previous Data}\\
\noalign{\smallskip}
\hline
\noalign{\smallskip}
Line	&Freq&(GHz)&Telescope &\multicolumn{4}{c}{Reference paper}\\ 
\noalign{\smallskip}
\hline
\noalign{\smallskip} 
HCN(4--3)&354&505 	&JCMT&\multicolumn{4}{c}{ Tacconi et al. (\cite{tacc94})}\\
CO(4--3) &461&041	&JCMT&\multicolumn{4}{c}{ Tacconi et al. (\cite{tacc94})}\\ 
\noalign{\smallskip}
\hline
\noalign{\smallskip}  
HCN(1--0)&89&088 	&IRAM PdBI&\multicolumn{4}{c}{ Tacconi et al. (\cite{tacc94})}\\
CS(2--1) &97&981 	&IRAM PdBI&\multicolumn{4}{c}{ Tacconi et al. (\cite{tacc97})}\\
CO(1--0) &115&271	&IRAM PdBI&\multicolumn{4}{c}{ Schinnerer et al. (\cite{schi00})}\\     
CO(2--1) &230&538	&IRAM PdBI&\multicolumn{4}{c}{ Schinnerer et al. (\cite{schi00})}\\  
\noalign{\smallskip}
\hline
\end{tabular}

%% file: n1068-usero-tab2.txt
\begin{tabular}{cclcr@{.}lcr@{.}lcr@{ }lcr@{ }l}
\hline
\hline
\noalign{\smallskip}
Position			&
$\;\;$  			&
{Line}  			&
$\;\;$  			&
\multicolumn{2}{c}{I(K km/s)}	&
$\;\;$  			&
\multicolumn{2}{c}{T$_{\mathrm{peak}}$(mK)}	&
$\;\;$  				   	&
\multicolumn{2}{c}{v-v$_{\mathrm{sys}}$(km/s)}	&
$\;\;$  					&
\multicolumn{2}{c}{$\Delta$v(km/s)}\\   
\noalign{\smallskip}
\hline
\noalign{\smallskip}
CND            & & SiO(2-1)              & & 0&56 (0.05) & & 2&8  & & -26&(10)  & & 189&(22)\\
(0\arcsec,0\arcsec)& & SiO(3-2)          & & 0&60 (0.06) & & 3&0  & & -36&(10)  & & 190&(19)\\
	       & & H$^{13}$CO$^{+}$(1-0) & & 0&57 (0.07) & & 2&1  & &   9&(13)  & & 254&(38)\\ 		
\noalign{\smallskip} 
\hline
\noalign{\smallskip}
S	       & & SiO(2-1)              & &  0&32 (0.04)& & 1&5  & & -48&(9)   & & 200&(13)\\
(0\arcsec,-16\arcsec)& & SiO(3-2)        & &  0&17 (0.04)& & 2&2  & & -57&(9)   & &  69&(18)\\
               & & H$^{13}$CO$^{+}$(1-0) & &  0&27 (0.04)& & 1&3  & & -48&(13)   & & 200&(13)\\
               & & HCO(3/2-1/2,2-1)      & &  0&11 (0.03)& & 1&1  & & -107&(13)  & & 100&(13)\\

\noalign{\smallskip} 
\hline
\noalign{\smallskip}
N	       & & SiO(2-1)              & & 0&39 (0.08) & & 1&4  & & -53&(29)  & &  261&(52)\\
(0\arcsec,+16\arcsec)& & SiO(3-2)	 & &$<$0&25      & & .&.  & & ...&      & &  ...&\\
               & & H$^{13}$CO$^{+}$(1-0) & & 0&44 (0.07) & & 2&8  & & -3&(11)   & &  145&(27)\\
               & & HCO(3/2-1/2,2-1)      & & 0&20 (0.07) & & 1&4  & & 26&(23)   & &  138&(60)\\

\noalign{\smallskip} 
\hline
\noalign{\smallskip}
E              & & SiO(2-1)              & & 0&28 (0.04) & & 1&8  & & -60&(19)   & &  150&(13)\\   
(+16\arcsec,0\arcsec)& & SiO(3-2)        & & 0&11 (0.04) & & 1&1  & & 1&(19)    & & 103&(36)\\  
               & & H$^{13}$CO$^{+}$(1-0) & & 0&21 (0.03) & & 2&2  & & -122&(9)  & &   90&(13)\\
               & & HCO(3/2-1/2,2-1)      & & 0&18 (0.04) & & 1&6  & & -130&(20)  & &   110&(13)\\

\noalign{\smallskip}
\hline
\end{tabular}

%% file: n1068-usero-tab3.txt
{
\begin{tabular}{lrrr}
\hline
\hline
\noalign{\smallskip}
Transition 		& I$_{\mathrm{blue}}$(K~km/s) & I$_{\mathrm{red}}$(K~km/s)& R$_{E/W}$\\ 
\noalign{\smallskip}
\hline
\noalign{\smallskip}
CO(1--0)		&127.1	(0.9)&	60.1 (0.9)&1.77 (0.03)    \\
CO(2--1)		&95.5	(0.4)&	51.2 (0.3)&1.56 (0.01)	\\
\noalign{\smallskip}
\hline
\noalign{\smallskip}
CS(2--1)		&8.7    (0.6)&  3.2 (0.5)&2.24 (0.40)	\\
\noalign{\smallskip}
\hline
\noalign{\smallskip}
HCN(1--0)		&121.5 (1.7) &  91.1 (1.5)&1.12 (0.02)	\\
\noalign{\smallskip}
\hline
\noalign{\smallskip}
SiO(2--1)		&8.5 (0.8)   &  6.2 (0.7)&1.15 (0.18)	\\
SiO(3--2)		&4.7 (0.5)   &  2.5 (0.4)&1.55 (0.32)	\\
\noalign{\smallskip}
\hline
\noalign{\smallskip}
H$^{13}$CO$^+$(1--0)	&6.3 (0.8)  &	6.9 (0.7)&0.77 (0.13)	\\
HCO$^+$(1--0)		&167.2 (2.2)&	158.4 (2.1)&0.88 (0.02)    \\
HOC$^+$(1--0)		&2.1 (0.5)  &	2.8 (0.4)& 0.61 (0.17)	\\

\noalign{\smallskip}
\hline
\end{tabular}
}

%% file: n1068-usero-tab4.txt
\begin{tabular}{lccl}
\hline
\hline
\noalign{\smallskip}
Transition 		&	$\langle${T}$\rangle_E$(K) 
& $\langle${T}$\rangle_W$(K)\\
\noalign{\smallskip}
\hline
\noalign{\smallskip}
CO(1--0)		&	4.34	&	2.44    \\
CO(2--1)		&	3.26	&	2.08	\\
CO(4--3)		&	7.58	&	6.50   \\
\noalign{\smallskip}
\hline
\noalign{\smallskip}
HCN(1--0)		&	4.15	&	3.71	\\
HCN(4--3)		&	1.24	&	0.62   \\
\noalign{\smallskip}
\hline
\noalign{\smallskip}
SiO(2--1)		&0.29		&	0.25	\\
SiO(3--2)		&0.16		&	0.10	\\
\noalign{\smallskip}
\hline
\noalign{\smallskip}
			&\multicolumn{2}{c}{T$_\mathrm{peak}$(K)}\\

CN(2--1, high freq.)	&\multicolumn{2}{c}{1.51}	\\
CN(2--1, low freq.) 	&\multicolumn{2}{c}{0.92}	\\
\noalign{\smallskip}
\hline
\noalign{\smallskip}
HCO$^+$(1--0)		&	5.73	&	6.50    \\
H$^{13}$CO$^+$(1--0)	&	0.21	&	0.28	\\
HOC$^+$(1--0)		&	0.07	&	0.12	\\
\noalign{\smallskip}
\hline
\noalign{\smallskip}
CS(2--1)		&	0.30	&	0.13	\\
\noalign{\smallskip}
\hline
\end{tabular}

%% file: n1068-usero-tab5.txt

\begin{tabular}{lllll}
\hline
\hline
\noalign{\smallskip}
species	& LVG-sol.    &	E/blue 		&W/red & N$_{\mathrm{E/W}}$\\ 
\noalign{\smallskip}
\hline
\noalign{\smallskip}
CO 
        & n 	      &	\dala{1.3}{4}	&\dala{2.5}{4} &   \\
	& N/$\Delta$v &	\dala{6.3}{15}	&\dala{4.0}{15}&1.0\\
	& X	      &	\dala{8.0}{-5}	&\dala{8.0}{-5}&   \\ 

\noalign{\smallskip}
\hline
\noalign{\smallskip}
SiO 
 	& n 	      &	\dala{1.6}{5} 	&\dala{6.0}{4} &   \\ 
	& N/$\Delta$v &	\dala{4.0}{11}	&\dala{5.0}{11}&0.5\\ 
        & X	      &	\dala{5.1}{-9} 	&\dala{1.0}{-8}&   \\ 
\noalign{\smallskip}
\hline
\noalign{\smallskip}
HCO$^+$ 
   	& n           &	 \dala{4.0}{4} 	&\dala{2.5}{4} &   \\ 
	& N/$\Delta$v &	 \dala{5.0}{12}	&\dala{8.0}{12}&0.4\\
	& X	      &	 \dala{6.3}{-8} &\dala{1.6}{-7}&   \\
\noalign{\smallskip}
\hline
\noalign{\smallskip}
HOC$^+$ 
 	& n           &	 ---		&\dala{2.5}{4} &   \\
 	& N/$\Delta$v &	 ---           	&\dala{2.0}{11}&---\\ 
        & X	      &	 ---            &\dala{4.0}{-9}&   \\
\noalign{\smallskip}
\hline
\noalign{\smallskip}
HCN 
 	& n           &	\dala{6.3}{5}	&\dala{4.0}{5} &   \\ 
        & N/$\Delta$v &	\dala{6.3}{12}	&\dala{6.3}{12}&0.8\\ 
        & X	      &	\dala{8.0}{-8}	&\dala{1.0}{-7}&   \\
\noalign{\smallskip}
\hline
\noalign{\smallskip}
CS
        & n           & \dala{6.3}{5}  	&\dala{4.0}{5} &   \\ 
	& N/$\Delta$v & \dala{1.6}{12} 	&\dala{5.0}{11}&1.2\\ 
	& X	      &	\dala{2.0}{-8} 	&\dala{1.6}{-8}&   \\
\noalign{\smallskip}
\hline
\noalign{\smallskip}
CN 	&	      &	\multicolumn{2}{c}{{\sl (global values)}}        &       \\
	& n &	\multicolumn{2}{c}{$>$\dala{1.6}{5} }            &	 \\
	& N/$\Delta$v &	\multicolumn{2}{c}{1$-$5\dala{}{13}}	 & ---   \\
	& X	      &	\multicolumn{2}{c}{\dala{9}{-8}$-$\dala{5}{-7}}&	 \\
\noalign{\smallskip}
\hline
\end{tabular}

%% file: n1068-usero-tab6.txt
\begin{tabular}{c r@{}l r@{}l r@{}l r@{}l r@{}l}
\hline
\hline
\noalign{\smallskip}
Abundance Ratios
&\multicolumn{2}{c}{CND(E/blue)}&\multicolumn{2}{c}{CND(W/red)}&\multicolumn{2}{c}{XDR} &\multicolumn{2}{c}{OER}	&OER+Ox&ygen Depletion\\  
\multicolumn{9}{c}{}	&\multicolumn{2}{c}{(Gas phase model)}\\ 
\noalign{\smallskip}
\hline
\noalign{\smallskip}
HCN/CO	   &1.0&$\times$10$^{-3}$  &1.2&$\times$10$^{-3}$  &5&$\times$10$^{-4}$
&1.0&$\times$10$^{-4}$& 1&$\times$10$^{-3}$\\ 
   
CS/CO      &2.5&$\times$10$^{-4}$ &1.6&$\times$10$^{-4}$&1-5&$\times$10$^{-4}$  &5.0&$\times$10$^{-5}$&
7&$\times$10$^{-5}$\\   
HCN/HCO$^+$&1.3&      &0.6& &0&.5--1&2.2&   & 25&\\ 

CN/HCN	   &\multicolumn{4}{c}{1--5 ({\sl global value})}& 3& &0.7&& 20&\\

SiO/CO	   &6.4&$\times$10$^{-5}$ &1.2&$\times$10$^{-4}$ &--&--& $<$6.6&$\times$10$^{-6}$&--&--\\ 
HCO$^+$/HOC$^+$& --&--& 50& & --&--& --&--& --&--\\  
\noalign{\smallskip}
\hline
\end{tabular}